\documentclass{article}
\usepackage[utf8]{inputenc}
\usepackage[margin=3cm]{geometry}
\usepackage{natbib}
\usepackage{amsmath, amssymb, graphicx, psfrag}
\usepackage{algorithm, algorithmicx}
\usepackage{algpseudocode}
\usepackage{mathtools}
\usepackage{changepage}
\usepackage{multirow}
\usepackage{authblk} 

\newcommand{\keywords}[1]{\textbf{\textit{Keywords---}} #1}

\DeclareMathAlphabet{\mymathbb}{U}{bbold}{m}{n}

\newcommand{\ceil}[1]{\left\lceil #1\right\rceil}
\newcommand{\floor}[1]{\left\lfloor #1\right\rfloor}

\makeatletter
\algnewcommand{\LineComment}[1]{{\Statex \hskip\ALG@thistlm \footnotesize\textcolor{blue}{/* #1 */}}}
\algnewcommand{\LineCommentIndent}[1]{{\Statex \hskip\ALG@tlm \footnotesize\textcolor{blue}{/* #1 */}}}
\makeatother

\title{Component-wise iterative ensemble Kalman inversion for static Bayesian models with unknown measurement error covariance}
\author[$\star$,$\diamondsuit$]{Imke Botha} 
\author[$\star$,$\dagger$,$\diamondsuit$]{Matthew P. Adams}
\author[$\star$]{Dang Khuong Tran}
\author[$\ddagger$]{Frederick R. Bennett}
\author[$\star$,$\diamondsuit$]{Christopher Drovandi}

\affil[$\star$]{School of Mathematical Sciences, Queensland University of Technology, Brisbane QLD, Australia}
\affil[$\dagger$]{School of Chemical Engineering, The University of Queensland, St Lucia QLD, Australia}
\affil[$\diamondsuit$]{QUT Centre for Data Science, Queensland University of Technology, Brisbane QLD, Australia}
\affil[$\ddagger$]{Department of Environment and Science, Queensland Government, Brisbane QLD Australia}

\date{\today}

\begin{document}

\setlength{\parindent}{0pc}
\setlength{\parskip}{1ex}

\maketitle
\noindent

\begin{abstract}
	The ensemble Kalman filter (EnKF) is a Monte Carlo approximation of the Kalman filter for high dimensional linear Gaussian state space models. EnKF methods have also been developed for parameter inference of static Bayesian models with a Gaussian likelihood, in a way that is analogous to likelihood tempering sequential Monte Carlo (SMC). These methods are commonly referred to as ensemble Kalman inversion (EKI). Unlike SMC, the inference from EKI is only asymptotically unbiased if the likelihood is linear Gaussian and the priors are Gaussian. However, EKI is significantly faster to run. Currently, a large limitation of EKI methods is that the covariance of the measurement error is assumed to be fully known. We develop a new method, which we call component-wise iterative ensemble Kalman inversion (CW-IEKI), that allows elements of the covariance matrix to be inferred alongside the model parameters at negligible extra cost. This novel method is compared to SMC on three different application examples: a model of nitrogen mineralisation in soil that is based on the Agricultural Production Systems Simulator (APSIM), a model predicting seagrass decline due to stress from water temperature and light, and a model predicting coral calcification rates. On all of these examples, we find that CW-IEKI has relatively similar predictive performance to SMC, albeit with greater uncertainty, and it has a significantly faster run time.
\end{abstract}


\section{Introduction} \label{sec:introduction}

Consider the following statistical model
\begin{align}
	y &= G(\theta) + \eta, \quad \eta \sim \mathcal{N}(0,\Gamma),
	\label{eq:static-model}
\end{align}
where $y \in Y \subseteq \mathbb{R}^{d_y\times 1}$ are the observations, $\theta \in \Theta \subseteq \mathbb{R}^{d_{\theta}\times 1}$ are the parameters, $G: \Theta \rightarrow Y$ is a deterministic mathematical model, and $\eta \in Y$ are the measurement errors. The number of observations is $d_y$, the number of parameters is $d_{\theta}$ and $\Gamma \in \mathbb{R}^{d_y\times d_y}$ is the covariance matrix characterising the measurement errors. Our interest is in the posterior distribution of the static parameters $\theta$ conditional on the observed data $y$,
\begin{align*}
	p(\theta|y) \propto \mathcal{N}(y\mid G(\theta), \Gamma)p(\theta),
\end{align*}
where $\mathcal{N}(y\mid G(\theta), \Gamma)$ is the likelihood function and $p(\theta)$ is the prior density of $\theta$.

Markov chain Monte Carlo \citep[MCMC;][]{Robert1999} or sequential Monte Carlo methods \citep[SMC;][]{DelMoral2006} can be used for asymptotically exact parameter inference of $\theta$ and $\Gamma$. These methods generally require many evaluations of $G(\cdot)$, however, which limits their feasibility when $G(\cdot)$ is computationally expensive to evaluate. 

If the elements of the covariance matrix $\Gamma$ are known, methods based on the ensemble Kalman filter \citep[EnKF;][]{Evensen1994, Burgers1998} are a fast but inexact alternative to MCMC and SMC. EnKF is a Monte Carlo approximation of the Kalman filter for state estimation of high dimensional linear Gaussian state space models (LG-SSMs). Unlike the Kalman filter, EnKF can also be applied to non-linear and non-Gaussian state space models, but the resulting inference is only asymptotically unbiased for LG-SSMs \citep{LeGland2009,Roth2017}.

\citet{Iglesias2013} extend EnKF for inverse problems, and the resulting methods are generally known as ensemble Kalman inversion (EKI). In the Bayesian setting, an initial ensemble is simulated from the prior, and iteratively updated to capture statistical properties of the EKI approximation $\widehat{p}(\theta\mid y)$ to the posterior $p(\theta\mid y)$. If the EKI algorithm is iterated long enough, i.e.\ as the number of iterations $J$ approaches infinity, the ensemble will collapse to a single point which is a minimiser of the loss function ${||y - G(\theta)||}_{\Gamma}$. For uncertainty quantification, i.e.\ to obtain samples from $\widehat{p}(\theta\mid y)$, early stopping of the algorithm is essential \citep{Iglesias2013}. If $G(\cdot)$ is a linear function and the prior is Gaussian, EKI provides exact samples from the posterior in a single iteration ($J = 1$) \citep{Iglesias2014,Duffield2021}, and the ensemble converges to the maximum a posteriori (MAP) estimate as $J \rightarrow \infty$ \citep{Iglesias2013,Duffield2021}. 

As inverse problems are often ill-posed, regularisation is required. For EKI, regularisation is induced by the subspace property, i.e.\ that the final ensemble is in the linear span of the initial ensemble \citep{Iglesias2013}. Additional regularisation can be used to improve the robustness and stability of EKI for sampling from $\widehat{p}(\theta\mid y)$, and to avoid overfitting the data. To that end, \citet{Iglesias2014} introduces an iteratively regularised extension of EKI. In the linear case, this method targets the power posterior $\pi_{j}(\theta) \propto p(y|\theta)^{\alpha_j}p(\theta)$, where $0 \le \alpha_j \le 1$ is the regularisation parameter at iteration $j$ and $\alpha_{j+1} > \alpha_{j}$ for $j = 0,\ldots, J-1$ \citep{Iglesias2018}. Other forms of regularisation can also be applied \citep{Chada2020}.

\citet{Iglesias2018} propose an adaptive version of the iterative EKI method of \citet{Iglesias2014}, which is analogous to density tempering SMC \citep{DelMoral2006}. An alternative adaptation method is given by \citet{Iglesias2021}. The ensemble Kalman sampler \citep{GarbunoInigo2020,Ding2021} is a variation of EKI, which perturbs the ensemble instead of the observations as in regular EKI. \Citet{Duffield2021} extend EKI for non-Gaussian likelihoods through a Gaussian approximation of the likelihood, and \citet{Wu2022} use EKI as the forward kernel in data annealing SMC. The latter method is exact, and requires much fewer evaluations of $G(\cdot)$ than SMC with a Metropolis-Hastings forward kernel. Also of note is the extension given by \citet{Rammay2020} to account for model misspecification.

Currently, a limiting factor of all these methods is that the noise $\eta$ is assumed to be characterised by known covariance matrix $\Gamma$. In this paper we develop a new adaptive iterative EKI method, which we call component-wise iterative ensemble Kalman inversion (CW-IEKI). This new method extends that of \citet{Iglesias2018}, and it can handle the common situation where $\Gamma$ contains unknown elements $\phi$ that require estimation. The method that we develop is completely analogous to density tempering SMC, which permits a direct comparison with the latter in terms of posterior accuracy and computation time. A comparison of a similar iterative EKI method \citep[following the method of ][]{Rammay2020} and density tempering SMC is provided in \citet{Vilas2021}; however, for the iterative EKI method they estimate $\phi$ as a function of the remaining parameters and they only consider one model example. In this paper we illustrate the new CW-IEKI method and perform comparisons on three model examples: a model of nitrogen mineralisation in soil that is based on the Agricultural Production Systems Simulator (APSIM) with partially known noise \citep{Vilas2021}, a model predicting seagrass decline due to cumulative stress from water temperature and light \citep{Adams2020}, and a model for predicting coral calcification rates \citep{Galli2018}.

The rest of the paper is organized as follows. Section \ref{sec:background} gives the background on EnKF, particle filters, likelihood tempering SMC, and iterative EKI methods. Section \ref{sec:methods} describes our novel CW-IEKI method and Section \ref{sec:examples} compares the performance of our method on three ecological model examples. Section \ref{sec:discussion} concludes.

\section{Background} \label{sec:background}

This section describes the filtering problem and the solutions given by the EnKF \citep{Evensen1994a} and the bootstrap particle filter \citep{Gordon1993}. It also describes how likelihood tempering SMC and EKI can be used for parameter inference of the static model given in equation \eqref{eq:static-model}. We use the notation ${z}_{i:j} \coloneqq \{z_i, z_{i+1}, \ldots, z_j\}^\top$ for $j\ge i$ throughout.

\subsection{Ensemble Kalman Filter}
The Kalman filter and the EnKF were developed to solve the filtering problem for state space models --- this is often referred to as state estimation or data assimilation. Consider a state space model of the form
\begin{align}
	\begin{split}
	Y_{t}\mid (X_{t} = x_{t}) &\sim g(y_t\mid x_{t}), \\
	X_{t}\mid (X_{t-1} = x_{t-1}) &\sim f(x_{t}\mid x_{t-1}),  \\
	X_0 &\sim \mu(x_0),
	\end{split}
	\label{eq:SSM}
\end{align}
where $y_{1:T}$ are the observed data, $x_{0:T}$ are the unobserved or latent states and $T$ is the number of observations. The filtering distribution, $p(x_t \mid y_{1:t})$, can be solved by recursively applying the time update 
\begin{align}
	p(x_t\mid y_{1:t-1}) = \int{f(x_t\mid x_{t-1})p(x_{t-1}\mid y_{1:t-1})}d{x_{t-1}}. \label{eqn:time_update}
\end{align}
and the measurement update
\begin{align}
	p(x_t \mid y_{1:t}) = \frac{g(y_t\mid x_t)p(x_t\mid y_{1:t-1})}{p(y_t\mid y_{t-1})}. \label{eqn:measurement_update}
\end{align}
Note that $p(x_0 \mid y_{1:0}) = \mu(x_0)$ is assumed to be known. See Chapter 3 of \citet{Schoen2017} for more detail on the filtering problem and its solution. The Kalman filter solves \eqref{eqn:time_update}-\eqref{eqn:measurement_update} analytically for LG-SSMs,
\begin{align*}
	\begin{split}
		Y_{t}\mid(X_{t} = x_{t}) &\sim \mathcal{N}(y_{t}\mid Hx_{t}, R), \\
		X_{t}\mid (X_{t-1} = x_{t-1}) &\sim \mathcal{N}(x_{t}\mid Fx_{t-1}, Q),  \\
		X_0 &\sim \mathcal{N}(x_0\mid\hat{x}_0, C_0^{xx}),
	\end{split}	
\end{align*}
where $y_t \subseteq \mathbb{R}^{d_y\times 1}$ and $x_t \subseteq \mathbb{R}^{d_x\times 1}$. The quantities $H \subseteq \mathbb{R}^{d_y\times d_x}$, $R \subseteq \mathbb{R}^{d_y\times d_y}$, $\hat{x}_0 \subseteq \mathbb{R}^{d_x\times 1}$ and $F, Q, C_0^{xx} \subseteq \mathbb{R}^{d_x\times d_x}$ are assumed to be known. Here, the time \eqref{eqn:time_update} and measurement \eqref{eqn:measurement_update} updates are 
\begin{align*}
	p(x_t\mid y_{1:t-1}) &= \mathcal{N}(x_t\mid \tilde{x}_{t}, C_{t}^{\tilde{x}\tilde{x}}), \\
	\tilde{x}_{t} &= F\hat{x}_{t-1}, \nonumber\\
	C_{t}^{\tilde{x}\tilde{x}} &= FC_{t-1}^{xx}F^\top + Q,
\end{align*}
and 
\begin{align*}
	p(x_t \mid y_{1:t}) &= \mathcal{N}(x_t\mid \hat{x}_{t}, C_{t}^{xx}), \\
	\hat{x}_{t} &= \tilde{x}_{t} + K_t(y_t - \tilde{y}_t), \\
	C_{t}^{xx} &= (I - K_tH)C_{t}^{\tilde{x}\tilde{x}},
\end{align*}
respectively, where $\tilde{y}_t = H \tilde{x}_{t}$ and $I\in \mathbb{R}^{d_x\times d_x}$ is the identity matrix. The Kalman gain at time $t$ is
\begin{align*}
	K_t = C_{t}^{\tilde{x}\tilde{y}}{(C_{t}^{\tilde{y}\tilde{y}})}^{-1} = C_{t}^{\tilde{x}\tilde{x}} H^{\top}(HC_{t}^{\tilde{x}\tilde{x}}H^\top + R)^{-1},
\end{align*}
where $C_{t}^{\tilde{x}\tilde{y}}$ is the cross covariance between $\tilde{x}_t$ and $\tilde{y}_t$, $C_{t}^{\tilde{y}\tilde{y}}$ is the covariance of $\tilde{y}_t$, and $C_{t}^{\tilde{x}\tilde{x}}$ is the covariance of $\tilde{x}_{t}$.

The EnKF is a Monte Carlo approximation of the Kalman filter for LG-SSMs. For non-linear, non-Gaussian SSMs, EnKF is asymptotically biased. The EnKF simulates the initial ensemble from the prior $\hat{x}_{0}^{n}\sim\mu(x_0)$ for $n = 1, \ldots, N$, then for each iteration, the ensemble is updated as follows. First, the state and observation predictions are simulated 
\begin{align*}
	\tilde{x}_t^{n} &\sim f(\cdot\mid \hat{x}_{t-1}^{n}), \\
	\tilde{y}_t^{n} &\sim g(\cdot\mid \tilde{x}_t^{n}),
\end{align*}
for $n = 1,\ldots, N$. In the linear case, $\hat{x}_{t-1}^{n} \sim p(x_{t-1} \mid y_{1:t-1})$, and $\tilde{x}_{t}^{n} \sim p(x_{t} \mid y_{1:t-1})$. Then, the ensemble is given by
\begin{align*}
	\hat{x}_{t}^{n} = \tilde{x}_{t}^{n} + \widehat{C}_t^{\hat{x}\tilde{y}}{(\widehat{C}_t^{\tilde{y}\tilde{y}})}^{-1} (y_t - \tilde{y}_{t}^{n}),
\end{align*}
where $\widehat{C}_t^{\hat{x}\tilde{y}}$ is the sample cross covariance between $\tilde{x}_t\in \mathbb{R}^{dx \times N}$ and $\tilde{y}_t\in \mathbb{R}^{dy \times N}$,
\begin{align}
	\widehat{C}_t^{\tilde{x}\tilde{y}} = \frac{1}{N-1}\sum_{n=1}^N{\left(\tilde{x}_t^{n} - \frac{1}{N}\sum_{j=1}^N{\tilde{x}_t^{j}}\right)\left(\tilde{y}_t^{n} - \frac{1}{N}\sum_{j=1}^N{\tilde{y}_t^{j}}\right)^{\top}}, \label{eq:Cxy}
\end{align}
and $\widehat{C}_t^{\tilde{y}\tilde{y}}$ is the sample covariance of $\tilde{y}_t\in \mathbb{R}^{dy \times N}$
\begin{align}
	\widehat{C}_t^{\tilde{y}\tilde{y}} = \frac{1}{N-1}\sum_{n=1}^N{\left(\tilde{y}_t^{n} - \frac{1}{N}\sum_{j=1}^N{\tilde{y}_t^{j}}\right)\left(\tilde{y}_t^{n} - \frac{1}{N}\sum_{j=1}^N{\tilde{y}_t^{j}}\right)^{\top}}.\label{eq:Cyy}
\end{align}
For LG-SSMs, $\hat{x}_{t}^{n} \sim p(x_{t} \mid y_{1:t})$. If the observation density is Gaussian, i.e.\ $g(y_t\mid x_t) = \mathcal{N}(y_t\mid G(x_t), \Gamma)$ with known covariance $\Gamma$, the Monte Carlo error in the calculation of the covariance matrices can be reduced \citep{Roth2017}. Let $\tilde{g}_t^n = G(\tilde{x}_t^n)$, then the sample cross covariance and sample covariance defined in equations \eqref{eq:Cxy} and \eqref{eq:Cyy} become
\begin{align}
	\widehat{C}_t^{\tilde{x}\tilde{y}} &= \frac{1}{N-1}\sum_{n = 1}^{N}\left(\tilde{x}_{t}^{n} - \frac{1}{N}\sum_{j=1}^N\tilde{x}_{t}^{j}\right)\left(\tilde{g}_{t}^{n} - \frac{1}{N}\sum_{j=1}^N\tilde{g}_{t}^{j}\right)^{\top} \label{eq:ccov_xy}, \\
	\widehat{C}_t^{\tilde{y}\tilde{y}} &= \frac{1}{N-1}\sum_{n = 1}^{N}\left(\tilde{g}_{t}^{n} - \frac{1}{N}\sum_{j=1}^N\tilde{g}_{t}^{j}\right)\left(\tilde{g}_{t}^{n} - \frac{1}{N}\sum_{j=1}^N\tilde{g}_{t}^{j}\right)^{\top} + \Gamma. \label{eq:ccov_yy}
\end{align}

\subsection{Particle Filters}

For non-linear, non-Gaussian state space models, sequential Monte Carlo (SMC) methods give an exact solution to the filtering problem. SMC methods for dynamic models are often referred to as particle filters. As with EnKF, the bootstrap particle filter \citep{Gordon1993} draws an initial ensemble from the prior $\mu(x_0)$. The particle filter then transforms the prior ensemble to samples from the filtering distribution through a sequence of reweighting, resampling and mutation steps. Given a set of weighted samples, $\{\hat{x}_{t-1}^n, W_{t-1}^n\}_{n=1}^{N} \sim p(x_{t-1} \mid y_{1:{t-1}})$, the bootstrap particle filter maps these to $p(x_{t} \mid y_{1:{t}})$ as follows:
\begin{enumerate}
	\item Resample the particles according to their weights, $W_{t-1}^{1:N}$ and set $W_{t-1}^n = 1\slash N$ for $n = 1, \ldots, N$. This gives a set of evenly weighted particles $\{\hat{x}_{t-1}^n, \frac{1}{N}\}_{n=1}^{N}$ distributed according to $p(x_{t-1} \mid y_{1:{t-1}})$.
	\item Simulate the state predictions using the transition density $\tilde{x}_t^n \sim f(\cdot\mid \hat{x}_{t-1}^n)$, which gives a set of unweighted particles distributed according to $p(x_{t} \mid y_{1:{t-1}})$.
	\item Reweight the particles using the observation density $w_t^n = g(y_t\mid \tilde{x}_t^n)$ for $n = 1,\ldots, N$ and normalise the weights to get $W_t^{1:N}$. The final set of weighted particles is distributed according to $p(x_{t} \mid y_{1:t})$.
\end{enumerate}
Steps 1-3 are iterated until all $T$ observations have been processed. When $t=0$, $p(x_{0} \mid y_{1:{0}})=\mu(x_0)$.

\subsection{Likelihood Tempering SMC} \label{sec:SMC}

SMC can also be used to sample from the static model given in equation \eqref{eq:static-model} \citep{DelMoral2006}. SMC methods require a sequence of distributions, $\pi_0(\theta), \ldots, \pi_J(\theta)$ to be defined, where $\pi_J(\theta)$ is equal to the desired posterior distribution $p(\theta\mid y)$. A common approach is likelihood tempering SMC, which raises the likelihood function to a power $\alpha_j, j= 0,\ldots, J$, where $\alpha_0 = 0 \le \alpha_1 \le \cdots \le \alpha_J = 1$. At iteration $j$, the power posterior $\pi_j(\theta) \propto p(y\mid\theta)^{\alpha_j}p(\theta)$ is targeted. Note that $\pi_0(\theta) = p(\theta)$ is the prior and $\pi_J(\theta) \propto p(y\mid\theta)p(\theta)$ is the posterior distribution. 

Given a set of evenly weighted samples from $\pi_{j-1}(\theta)$, likelihood tempering SMC transforms these to samples from $\pi_j(\theta)$ as follows:
\begin{enumerate}
	\item Reweight the particles using the ratio of the current target to the previous target, $w_j^n = \pi_j(\theta_{j-1}^n)\slash \pi_{j-1}(\theta_{j-1}^n)$ for $n = 1,\ldots, N$ and normalise the weights to get $W_j^{1:N}$. This gives a set of weighted particles that are distributed according to $\pi_j(\theta)$. 
	\item Resample the particles according to their weights, and set $W_{j-1}^n = 1\slash N$ for $n = 1, \ldots, N$. 
	\item Mutate the resampled particles using a Markov chain Monte Carlo (MCMC) kernel which targets the distribution $\pi_j(\theta)$.
\end{enumerate}
Step 2 removes the negligible weight particles and duplicates the high weight particles, and Step 3 diversifies the particles to mitigate the duplication. A common approach to mutate the particles is to use $M$ iterations of an MCMC algorithm with $\pi_j(\theta)$ as its invariant distribution. 

The tempering parameter $\alpha_j$ can be adapted at each iteration by setting $\alpha_j$ such that a pre-specified effective sample size (ESS) threshold is achieved \citep{Jasra2010}. This will be some proportion of $N$. While the ESS cannot be calculated exactly, it can be approximated at each iteration $j$ using the normalised weights $W_j^{1:N}$, 
\begin{align}
	\textrm{ESS}_j = \frac{1}{\sum_{n=1}^N(W_j^n)^2}.
	\label{eq:ESS}
\end{align} 
Likelihood tempering SMC can also be used if the noise parameter $\Gamma$ has unknown elements $\phi$. In this case, the method is applied to $\{\theta, \phi\}$ instead of $\theta$. 

If the function $G(\cdot)$ in equation \eqref{eq:static-model} is expensive to compute, SMC may be prohibitively expensive to run. Each iteration $j = 1, \ldots, J$ requires a minimum of $NM_j$ evaluations of $G(\cdot)$, where $M_j$ is the number of MCMC repeats in iteration $j$. The entire algorithm requires a minimum of $N\sum_{j = 1}^{J}{M_j} + 1$ evaluations, where the extra evaluation comes from the initial calculation of the likelihood. A less expensive, but asymptotically biased alternative to SMC for static models is ensemble Kalman inversion.  

\subsection{Ensemble Kalman Inversion}

\citet{Iglesias2013} extend the EnKF algorithm for static models with known $\Gamma$ by introducing artificial dynamics. The static model in equation \eqref{eq:static-model} can be constructed from the general state space model in equation \eqref{eq:SSM} by setting the transition density to the identity function and denoting $x_{t} = \theta_j$, i.e.\ $f(x_{t}\mid x_{t-1}) = x_{t-1} = \theta_{j-1}$. The EnKF artificial time update is then $\tilde{\theta}_j^n = \theta_{j-1}^{n}$ for $n = 1,\ldots, N$ and the measurement update is $\theta_j^n = \tilde{\theta}_j^n + \hat{C}_{j}^{\tilde{\theta}\tilde{y}}(\hat{C}_{j}^{\tilde{y}\tilde{y}})^{-1} (y_j - \tilde{y}_{j}^n)$, where $\tilde{y}_j^n\sim \mathcal{N}(\cdot\mid G(\tilde{\theta}_{j}^n), \Gamma)$. The EKI algorithm of \citet{Iglesias2013} for static Bayesian models proceeds as below:
\begin{enumerate}
	\item Sample $\theta_0^{n} \sim p(\theta)$ for $n = 1,\ldots,N$.
	\item Update $\theta_j^{n} = \tilde{\theta}_j^n + \hat{C}_{j}^{\tilde{\theta}\tilde{y}}(\hat{C}_{j}^{\tilde{y}\tilde{y}})^{-1}(y - \tilde{y}_{j}^n)$ where $\tilde{\theta}_j^n = \theta_{j-1}^{n}$ and $\tilde{y}_{j}^n \sim \mathcal{N}(G(\tilde{\theta}_j^n),\Gamma)$ for $n = 1,\ldots,N$.
	\item Iterate Step 2 as desired.
\end{enumerate}
Since the likelihood $p(y\mid\theta)$ is Gaussian, equations \eqref{eq:ccov_xy} and \eqref{eq:ccov_yy} are used for the covariance calculations. In equation \eqref{eq:ccov_xy}, the ensemble $\tilde{x}_t \in \mathbb{R}^{d_x\times N}$ is replaced with $\tilde{\theta}_j \in \mathbb{R}^{d_{\theta}\times N}$.

While the prior induces regularisation through the subspace property, additional regularisation is often required to properly explore regions of high posterior support without overfitting the data \citep{Iglesias2014}. An iteratively regularised extension of the EKI method of \citet{Iglesias2013} is the algorithm of \citet{Iglesias2014}:
\begin{enumerate}
	\item Sample $\theta_0^{n} \sim p(\theta)$ for $n = 1,\ldots,N$.
	\item Update $\theta_j^{n} = \tilde{\theta}_j^n + \hat{C}_{j}^{\tilde{\theta}\tilde{y}}(\hat{C}_{j}^{\tilde{y}\tilde{y}})^{-1}(y - \tilde{y}_{j}^n)$ where $\tilde{\theta}_j^n = \theta_{j-1}^{n}$ and $\tilde{y}_{j}^n \sim \mathcal{N}(G(\tilde{\theta}_j^n),h_j^{-1}\Gamma)$ for $n = 1,\ldots,N$.
	\item Iterate Step 2 until $\sum_{i=1}^Jh_i = 1$.
\end{enumerate}
Here, \eqref{eq:ccov_yy} becomes
\begin{align*}
	\hat{C}_{j}^{\tilde{y}\tilde{y}} = \frac{1}{N-1}\sum_{n = 1}^{N}\left(\tilde{g}_{j}^n - \frac{1}{N}\sum_{k=1}^N\tilde{g}_{j}^k\right)\left(\tilde{g}_{j}^n - \frac{1}{N}\sum_{k=1}^N\tilde{g}_{j}^k\right)^{\top} + h_j^{-1}\Gamma.
\end{align*}

We refer to this method as iterative EKI (IEKI). Similar to likelihood tempering SMC, IEKI also targets a sequence of distributions $\tilde\pi_0(\theta), \ldots, \tilde\pi_J(\theta)$. At iteration $j$, the IEKI algorithm targets $\tilde\pi_j(\theta)$, which is an approximation to the power posterior 
\begin{align*}
	\pi_j(\theta) \propto \mathcal{N}(y\mid G(\theta),\Gamma)^{\alpha_j}p(\theta).
\end{align*}
For a linear model with a Gaussian prior, $\tilde\pi_j(\theta) = \pi_j(\theta)$. Note that $\pi_j(\theta)$ is exactly the $j$th target in the likelihood tempering SMC algorithm defined in Section \ref{sec:SMC}. 

The parameter $h_j$ for $j=1,\ldots, J$ can be chosen adaptively using the method of \citet{Iglesias2018}. At iteration $j$, assume that the particles must be reweighted from $\tilde\pi_{j-1}(\theta)$ to $\tilde\pi_j(\theta)$. Analogously to likelihood tempering SMC, these weights are given by 
\begin{align*}
	w_j^n = \frac{\pi_{j}(\theta_{j-1}^n)}{\pi_{j-1}(\theta_{j-1}^n)} \propto \exp{\left(-\frac{1}{2}h_j\left(y-G(\theta_{j-1}^n)^\top\Gamma^{-1}\left(y-G(\theta_{j-1}^n)\right)\right)\right)},
\end{align*}
where $h_j = \alpha_{j} - \alpha_{j-1}$, and the obtained $w_j^{1:N}$ are thereafter normalised to give $W_j^{1:N}$. The parameter $\alpha_j$ can be set so that the ESS, estimated using \eqref{eq:ESS}, matches some target threshold. Once $\alpha_j$ is chosen, $h_j$ is given by $\alpha_j - \sum_{i=1}^{J-1}h_i$. 

The function $G(\cdot)$ is evaluated once per particle at every iteration, so that the total number of evaluations for IEKI is $JN$, where $J$ is the total number of iterations and $N$ is the number of particles or the ensemble size. This is much less than the computation required for SMC, but the IEKI assumes that $\Gamma$ is known.  In the next section we develop a new adaptive IEKI method that can estimate unknown parameters associated with $\Gamma$.

\section{Component-Wise Iterative Ensemble Kalman Inversion} \label{sec:methods}

A strong limitation of IEKI is that $\Gamma$ must be known. We extend IEKI to the case where $\Gamma$ depends on some unknown parameter or parameters $\phi$. For example, in the simplest case, this might be $\Gamma(\phi) = \phi I$, where $I\in \mathbb{R}^{d_y\times d_y}$ is the identity matrix, although our method does not require this assumption to hold. The target distribution at iteration $j$ is $\tilde\pi_j(\theta,\phi)$, which approximates the power posterior
\begin{align*}
	\pi_j(\theta,\phi) \propto \mathcal{N}\left(y\mid G(\theta),\Gamma(\phi)\right)^{\alpha_j}p(\theta,\phi).
\end{align*} 
At each iteration $j = 1,\ldots, J$, the model parameters $\theta$ and the noise parameters $\phi$ are updated component-wise conditional on the other. We refer to our method as component-wise IEKI (CW-IEKI). Our proposed procedure for CW-IEKI is as follows:
\begin{enumerate}
	\item Sample $\{\theta_0^{n},\phi_0^{n}\} \sim p(\theta,\phi)$ for $n = 1,\ldots,N$.
	\item Update the model parameters $\theta$: $\theta_j^{n} = \tilde{\theta}_j^n + \hat{C}_{j}^{\tilde{\theta} \tilde{y}}\left(\hat{C}_{j}^{\tilde{y}\tilde{y}}(\phi_{j-1}^{n})\right)^{-1}\left(y - \tilde{y}_{j}^n\right)$ where $\tilde{\theta}_j^n = \theta_{j-1}^{n}$ and $\tilde{y}_{j}^n \sim \mathcal{N}(G(\tilde{\theta}_j^n),h_j^{-1}\Gamma(\phi_{j-1}^{n}))$ for $n = 1,\ldots,N$.
	\item Update the noise parameters $\phi$:  update $\phi_j^{n}$ conditional on  $\theta_j^{n}$ for $n = 1,\ldots,N$ using the Metropolis-Hastings MCMC update shown in Algorithm \ref{alg:noise_update}. 
	\item Iterate Steps 2 and 3 until $\sum_{i=1}^nh_i = 1$.
\end{enumerate}  

The covariance \eqref{eq:ccov_yy} in Step 2 is 
\begin{align*}
	\hat{C}_{j}^{\tilde{y}\tilde{y}}(\phi_{j-1}^{n}) = \frac{1}{N-1}\sum_{n = 1}^{N}\left(\tilde{g}_{j}^n - \frac{1}{N}\sum_{k=1}^N\tilde{g}_{j}^k\right)\left(\tilde{g}_{j}^n - \frac{1}{N}\sum_{k=1}^N\tilde{g}_{j}^k\right)^{\top} + h_j^{-1}\Gamma(\phi_{j-1}^{n}).
\end{align*}

In Step 3, the noise parameters $\phi$ are updated from the exact conditional posterior $\pi_j(\phi\mid\theta)$. We propose to use $M$ iterations of a Metropolis-Hastings MCMC kernel, where the ensemble $\phi_{j-1}^{1:N}$ can be used to inform the proposal distribution for $\phi$. If it is possible to independently sample from $\pi_j(\phi\mid\theta)$, then Gibbs sampling can also be used to update $\phi_{j-1}^{1:N}$. Note that Step 3 does not require evaluation of $G(\cdot)$ since $\theta$ is fixed. Consequently, the total number of evaluations of $G(\cdot)$ for our method is the same as for standard IEKI, i.e.\ $JN$, which again is typically much less than the $N\sum_{j = 1}^{J}{M_j} + 1$ evaluations required for likelihood tempering SMC. See Algorithm \ref{alg:noise_update} for more details. 

To adapt $h_j$, the weights are calculated in a similar way to IEKI,
\begin{align*}
	w_j^{n} &= \frac{\pi_j(\phi_{j-1}^n,\theta_{j-1}^n)}{\pi_{j-1}(\phi_{j-1}^n,\theta_{j-1}^n)} \\
	&\propto \exp \left(\log(h_j) - \frac{1}{2} \log \det \Gamma(\phi_{j-1}^{n})  -\frac{1}{2}h_j \left(y - G(\theta_{j-1}^{n})\right)^\top \Gamma(\phi_{j-1}^{n})^{-1} \left(y - G(\theta_{j-1}^{n})\right)  \right).
\end{align*}

Note that the likelihood covariance $\Gamma$ does not uniquely define the measurement error of the data in CW-IEKI as it does for standard IEKI. Since elements of $\Gamma$ are estimated, it may also capture aspects arising from model misspecification.  

\begin{algorithm}[htp]
	\begin{adjustwidth}{\algorithmicindent}{}
		\textbf{Input: } data $y$, ensembles $\theta_{j}^{1:N}$ and $\phi_{j-1}^{1:N}$, model evaluations $g_{j}^{n} = G(\theta_{j}^{n})$ for all $n=1,\ldots, N$ and $\alpha_j$ \\
		\textbf{Output: } updated ensemble of noise parameters $\phi_{j}^{1:N}$
	\end{adjustwidth}
	\vspace{0.5em}
	\begin{algorithmic}		
		\State Set $\phi_{j}^{1:N} = \phi_{j-1}^{1:N}$
		\For{$m=1$ to $M$}
		\For{$n = 1$ to $N$}
		\State Sample $\phi_{j}^{n, *}\sim q(\cdot\mid\phi_{j}^{n})$
		\vspace{0.5em}
		\State Calculate the acceptance probability
		\begin{align*}
			\alpha(\phi_{j}^{n}, \phi_{j}^{n, *}) = \min{\left(1, \frac{\mathcal{N}\left(y\mid g_{j}^{n},\Gamma(\phi_{j}^{n,*})\right)^{\alpha_j}p(\theta_{j}^{n}, \phi_{j}^{n, *})}{\mathcal{N}\left(y\mid g_{j}^{n},\Gamma(\phi_{j}^{n})\right)^{\alpha_j}p(\theta_{j}^{n}, \phi_{j}^{n})} \frac{q(\phi_{j}^{n}\mid\phi_{j}^{n, *})}{q(\phi_{j}^{n, *}\mid\phi_{j}^{n})}\right)}
		\end{align*}
		\vspace{0.5em}
		\State Sample $u\sim\operatorname{Uniform}(0, 1)$
		\vspace{0.5em}
		\If{$\alpha(\phi_{j}^{n}, \phi_{j}^{n, *}) < u$} 
			\State Set $\phi_{j}^{n} = \phi_{j}^{n, *}$
		\EndIf
		\EndFor
		\EndFor
	\end{algorithmic}
	\caption{MCMC update of the noise parameters $\phi$.}
	\label{alg:noise_update}
\end{algorithm}

\section{Performance of CW-IEKI} \label{sec:examples}

\subsection{Implementation of CW-IEKI and Likelihood Tempering SMC}

We compare our novel CW-IEKI method to likelihood tempering SMC on three model examples. The first is a model of nitrogen mineralisation in soil \citep{Vilas2021} that has relatively few parameters. The second model predicts seagrass decline due to cumulative water temperature and light stress \citep{Adams2020}, and the final model predicts coral calcification rates \citep{Galli2018}. The seagrass model has more parameters than the first model, and its marginal posteriors are roughly Gaussian. The coral model also has a relatively large number of parameters, but relatively uninformative data --- the marginal posteriors of the parameters are close to the priors.

All code is implemented in MATLAB. For both CW-IEKI and SMC, the ensemble size is fixed at $1000$, and the tempering schedule is adapted to achieve a target ESS of $N\slash 2 = 500$ unless otherwise specified. To mutate the noise ensemble ($\phi_{j}^{1:N}$) in CW-IEKI and the particles ($\{\theta, \phi\}_{j}^{1:N}$) in SMC we use a random walk Metropolis-Hastings kernel \citep{Hastings1970}, where the covariance of the random walk proposal is set to the covariance of the samples being mutated, i.e.\ $\textrm{cov}(\phi_{j}^{1:N})$ for CW-IEKI and $\textrm{cov}(\{\theta, \phi\}_{j}^{1:N})$ for SMC. Due to the higher number of parameters for the seagrass and coral models, the covariance of the random walk is scaled by $2.38^2\slash (d_{\theta} + d_{\phi})$ for SMC, where $d_{\theta}$ is the number of parameters in $\theta$ and $d_{\phi}$ is the number of parameters in $\phi$ \citep{Roberts2001}. For CW-IEKI, the number of MCMC iterations is fixed at a conservative $1000$ --- as no extra model evaluations are required, the cost of these iterations is relatively small. For SMC, the number of MCMC iterations is adapted at each iteration $j = 1, \ldots, J$ as follows \citep{South2019}:
\begin{enumerate}
	\item Run $S_j$ MCMC iterations and estimate the acceptance rate $p$.
	\item Adapt the total number of MCMC iterations as $M_j = \ceil{\log{(c)}\slash\log{(1-p)}}$.
	\item Complete the remaining $M_j - S_j$ MCMC iterations.
	\item Calculate $S_{j+1}$ for the next iteration as $S_{j+1} = \floor{M_j\slash 2}$.
\end{enumerate}
The value $1-c$ is the target acceptance rate, $\ceil{\cdot}$ denotes the ceiling function and $\floor{\cdot}$ denotes the floor function. For all models, $S_1 = 5$ and the target acceptance rate is $1 - 0.01 = 0.99$.

We assess the performance of CW-IEKI based on its accuracy, predictive performance and computation time relative to SMC. The marginal posterior density plots of the model parameters are used to compare the accuracy of CW-IEKI to the SMC solution. As these plots do not account for parameter interdependencies however, we also compare the marginal densities of parameter combinations that greatly influence the model fit \citep{MonsalveBravo2022}. These combinations are identified through the eigendecomposition of a sensitivity matrix that captures key characteristics of the posterior distribution. Unless otherwise specified, we calculate the sensitivity matrix as the inverse of the sample covariance of the natural logarithm of the posterior samples from SMC \citep{MonsalveBravo2022}. The logarithm of the $k$th parameter combination is
\begin{align}
	\alpha_k = \sum_{j = 1}^{d_{\theta}}{(v_k)_j\log{\theta_j}}, \label{eq:eigenparameters}
\end{align}
where $(v_k)_j$ is the $j$th element of the $k$th normalised eigenvector, and $\theta_j$ is the $j$th parameter. Following the terminology of \citet{MonsalveBravo2022}, we refer to \eqref{eq:eigenparameters} as the logarithm of the $k$th eigenparameter. The stiffest and sloppiest eigenparameters are those associated with the highest and lowest eigenvalues respectively. For all examples, the noise parameters $\phi$ are treated as nuisance parameters in the analysis of model sloppiness and are excluded when calculating the sensitivity matrix \citep{MonsalveBravo2022}.

Posterior predictive plots are used to assess the predictive performance of CW-IEKI relative to SMC. The posterior predictive distribution is given by
\begin{align*}
	p(y^*\mid y_{1:T}) = \int_{\Theta}{p(y^*\mid\theta)p(\theta\mid y_{1:T})}d\theta,
\end{align*}
which can be sampled by first sampling from the posterior distribution $\{\theta, \phi\}_J^* \sim p(\theta, \phi\mid y_{1:T})$, then sampling from the likelihood $y^*\mid \{\theta, \phi\}_J^* \sim \mathcal{N}(G(\theta_J^*), \Gamma(\phi_J^*))$. We compare the posterior predictive distribution estimated using the biased posterior samples from CW-IEKI to the posterior predictive distribution using the asymptotically exact samples from SMC. 

Since SMC is asymptotically unbiased, it is always expected to outperform CW-IEKI in terms of accuracy and predictive performance. The main advantage of CW-IEKI is a significant speed-up in computation time compared to SMC. We assume that the expense of evaluating the function $G(\cdot)$, i.e.\ the deterministic mean of the likelihood function, dominates the computation time. SMC has $N\sum_{j = 1}^{J}{M_j} + 1$ evaluations of $G(\cdot)$, while CW-IEKI only has $NJ$. The value of $N$ is fixed for both methods, while $J$ and $M_j, j = 1, \ldots, J$ are adapted. In general, $\sum_{j=1}^{J}{M_j} + 1 \gg J$.

\subsection{Model Example 1: Predicting Nitrogen Mineralisation} \label{sec:apsim}

The first model predicts cumulative nitrogen mineralisation, assuming a measurement error distributed according to \citep{Vilas2021}:
\begin{align*}
	y_{t_j}^r \sim \mathcal{N}\left(x_{t_j}, {(\zeta_{t_j}^r)}^2 + \sigma^2\right),  \quad x_{t_j} = G(\theta, {t_j}),
\end{align*}
for $j=1, \ldots, T$ and $r = 1, \ldots, R$, where $T$ is the number of timepoints, $R$ is the number of replicates per timepoint, and $x_{t_1}, \ldots, x_{t_T}$ are deterministic predictions of cumulative nitrogen mineralisation from version 7.10 of the APSIM model \citep{Holzworth2014} configured with soil water and nitrogen modules \citep{Probert1998}. The function $G(\cdot)$ has numerous parameters, most of which are fixed at measured values \citep{Probert1998}, apart from the model parameters we seek to obtain improved estimates for. Following the approach of \citet{Rammay2020}, the model error is separated into two parts, where the first term ($\zeta_{t_j}^r$) is known and accounts for measurement error, and the second term ($\sigma$) is unknown and accounts for all other sources of error such as model misspecification. At each timepoint $t_j, j=1,\ldots,T$ and replicate $r = 1, \ldots, R$, $\zeta_{t_j}^r$ is set to $4$\% of the observation $y_{t_j}^r$ \citep{APHA2012}.

We consider two versions of this model. The first estimates three parameters (fbiom, finert, $\sigma$) and is the one considered in \citet{Vilas2021}. For the second model, three additional model parameters are estimated (ef\_biom = ef\_hum, rd\_biom and rd\_hum) --- in the first model these parameters are fixed at ef\_biom = ef\_hum = $0.4$, rd\_biom = $0.0081$ and rd\_hum = $0.00015$. As a shorthand, in the present work we refer to these two models as the three parameter and six parameter APSIM models respectively. See \citet{Probert1998} for more detail about the model parameters and the values of the remaining parameters. The models are applied to data from \citet{Allen2019} measuring changes in inorganic nitrogen in soil from the Mackay Whitsundays region of North Queensland. The data is obtained from four $301$ day laboratory incubations (i.e.\ $R = 4$). The second model is also fitted to a dataset simulated using $\theta = \{\textrm{fbiom}, \textrm{finert}, \textrm{ef\_biom = ef\_hum}, \textrm{rd\_biom}, \textrm{rd\_hum}\} = \{0.1, 0.6, 0.3, 0.0025, 0.0005\}$ and $\phi=\sigma=8$. To enable simulation from the model, the known portion of the error ($\zeta_{t_j}^r$) is set to $4$\% of the mean at time $t_j$, i.e. for the synthetic dataset, $\zeta_{t_j}^r = 0.04\cdot x_{t_j}$ for all $r = 1, \ldots, 4$, matching the number of replicates in the data from \citet{Allen2019}.

We denote the truncated univariate normal distribution as $\mathcal{N}(x\mid \mu, \sigma^2, a, b)$, where $\mu$ is the mean, $\sigma$ is the standard deviation, $a$ is the lower bound, and $b$ is the upper bound. The assumed priors for fbiom, finert and $\sigma$ are $\mathcal{N}(\textrm{fbiom}\mid 0.093, 0.025^2, 0.05, 0.15)$, $\mathcal{N}(\textrm{finert}\mid 0.58, 0.1^2, 0.4, 0.8)$ and $\operatorname{Uniform}(\sigma\mid 0,20)$ for both models. For the second model, the additional priors are $\operatorname{Uniform}(\textrm{ef\_biom} = \textrm{ef\_hum}\mid 0,1)$, $\operatorname{Uniform}(\textrm{rd\_biom}\mid 0.001,0.01)$ and $\operatorname{Uniform}(\textrm{rd\_hum}\mid 0,0.001)$.

\subsubsection*{Three parameter APSIM model applied to the real data}

\begin{figure}[htp]
	\centering
	\includegraphics[scale=0.6]{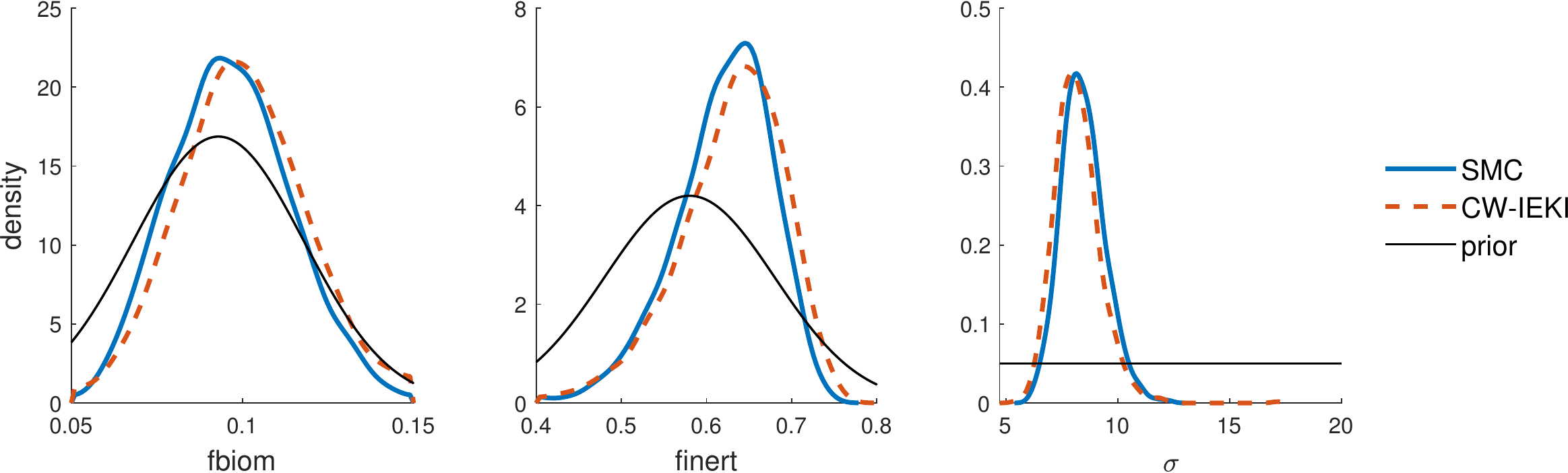}
	\caption{Marginal posterior density plots for the three parameter APSIM model applied to the real data.}
	\label{fig:2Params_densities}
\end{figure}

\begin{figure}[htp]
	\centering
	\includegraphics[scale=0.6]{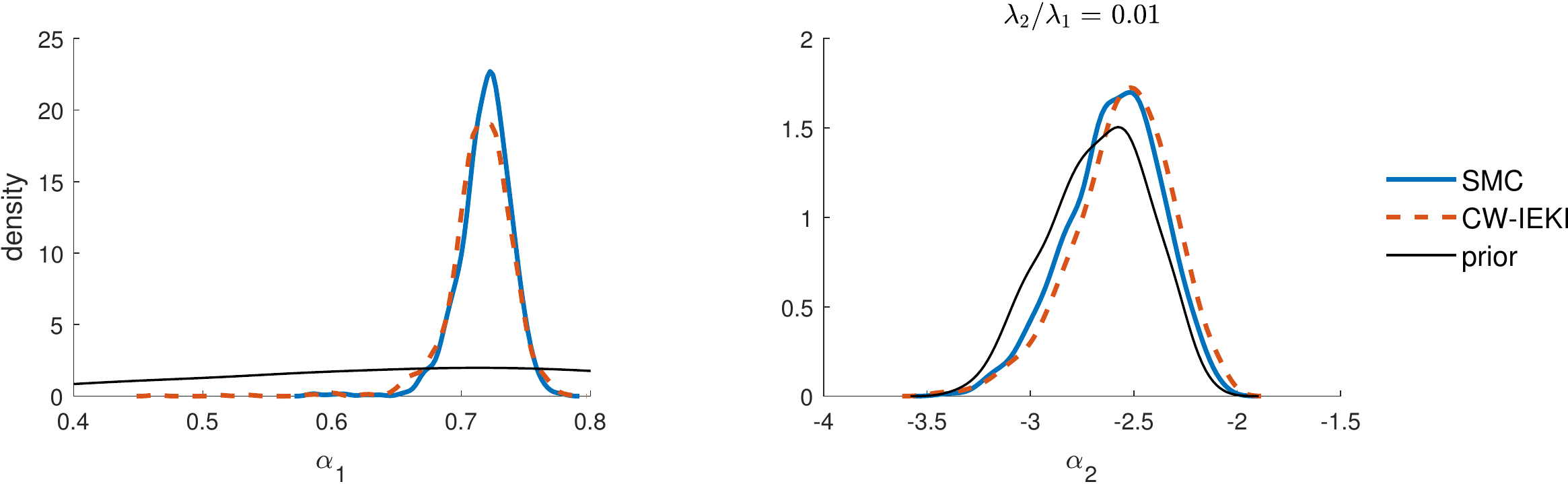}
	\caption{Marginal posterior density plots of the natural logarithm of the eigenparameters for the three parameter APSIM model applied to the real data. Note that the uncertainty parameter $\sigma$ is excluded from the analysis of sloppiness, and $\lambda_k$ is the eigenvalue associated with eigenvector $v_k$ in equation \eqref{eq:eigenparameters}. The logarithm of the eigenparameters are calculated based on samples from the prior (black), CW-IEKI (dashed red-orange) and SMC (blue) using a sensitivity matrix calculated using the SMC samples.}
	\label{fig:2Params_densities_eigenparameters}
\end{figure}

\begin{figure}[htp]
	\centering
	\includegraphics[scale=0.6]{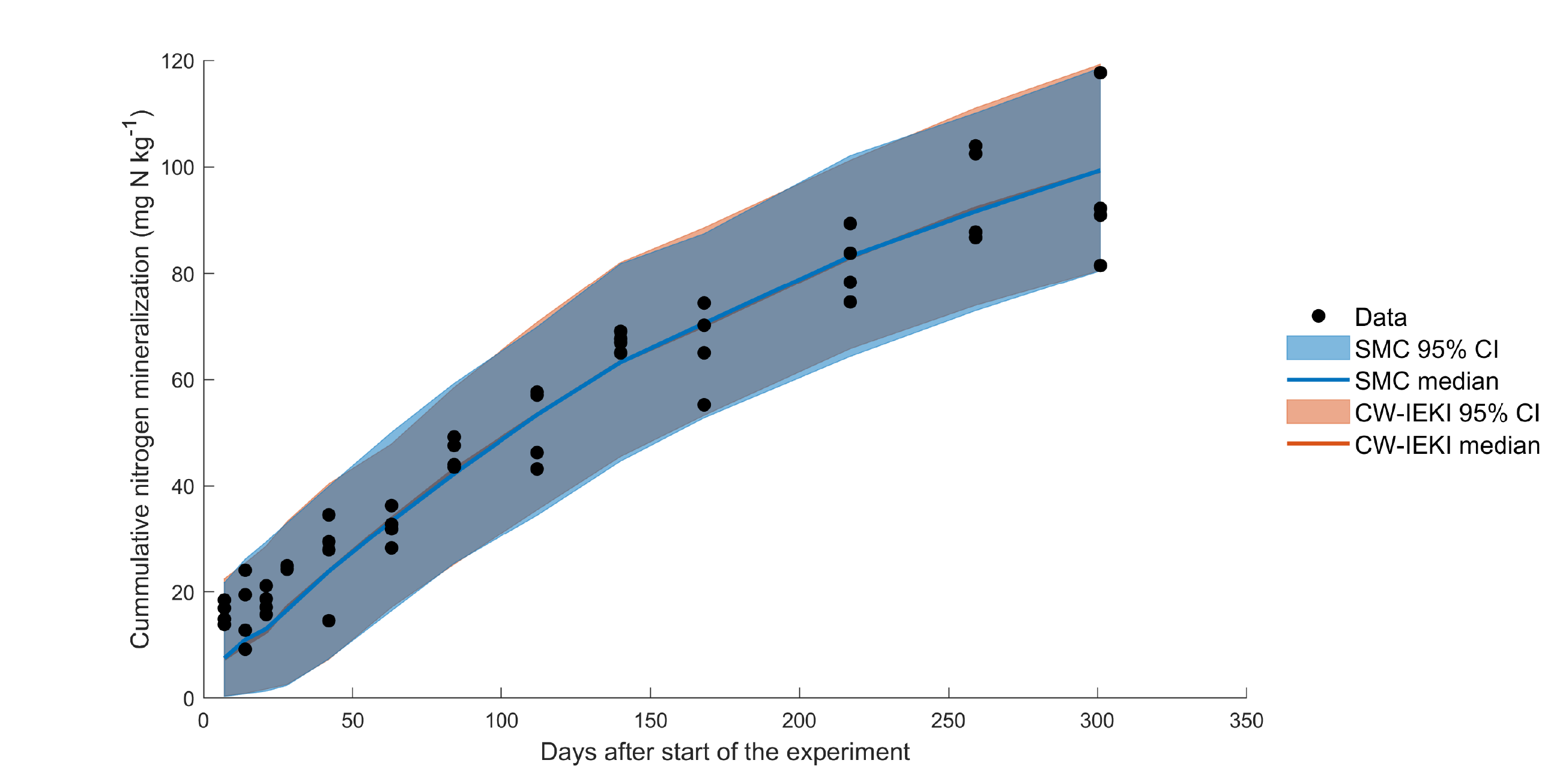}
	\caption{Comparison of the real data to the median and $95\%$ central credible intervals for the posterior predictive distribution of cumulative nitrogen mineralisation obtained from the three parameter APSIM model fitted to this data. Models were fitted using CW-IEKI (red-orange) and likelihood tempering SMC (blue).}
	\label{fig:2Params_postpred}
\end{figure}

Figures \ref{fig:2Params_densities} and \ref{fig:2Params_densities_eigenparameters} show the marginal posterior densities of the parameters and the eigenparameters of the three parameter APSIM model applied to the real data. Figure \ref{fig:2Params_postpred} shows the posterior predictive densities using CW-IEKI and SMC. On this example, both CW-IEKI and SMC have very similar results for accuracy and predictive performance. However, CW-IEKI is almost $11$ times faster than SMC with $5000$ evaluations of $G(\cdot)$ compared to $54000$ for SMC. (The number of evaluations of $G(\cdot)$ in our study is always a multiple of $1000$ because our chosen ensemble sizes for both CW-IEKI and SMC are $N = 1000$.)

\subsubsection*{Six parameter APSIM model applied to the real data}

\begin{figure}[htp]
	\centering
	\includegraphics[scale=0.6]{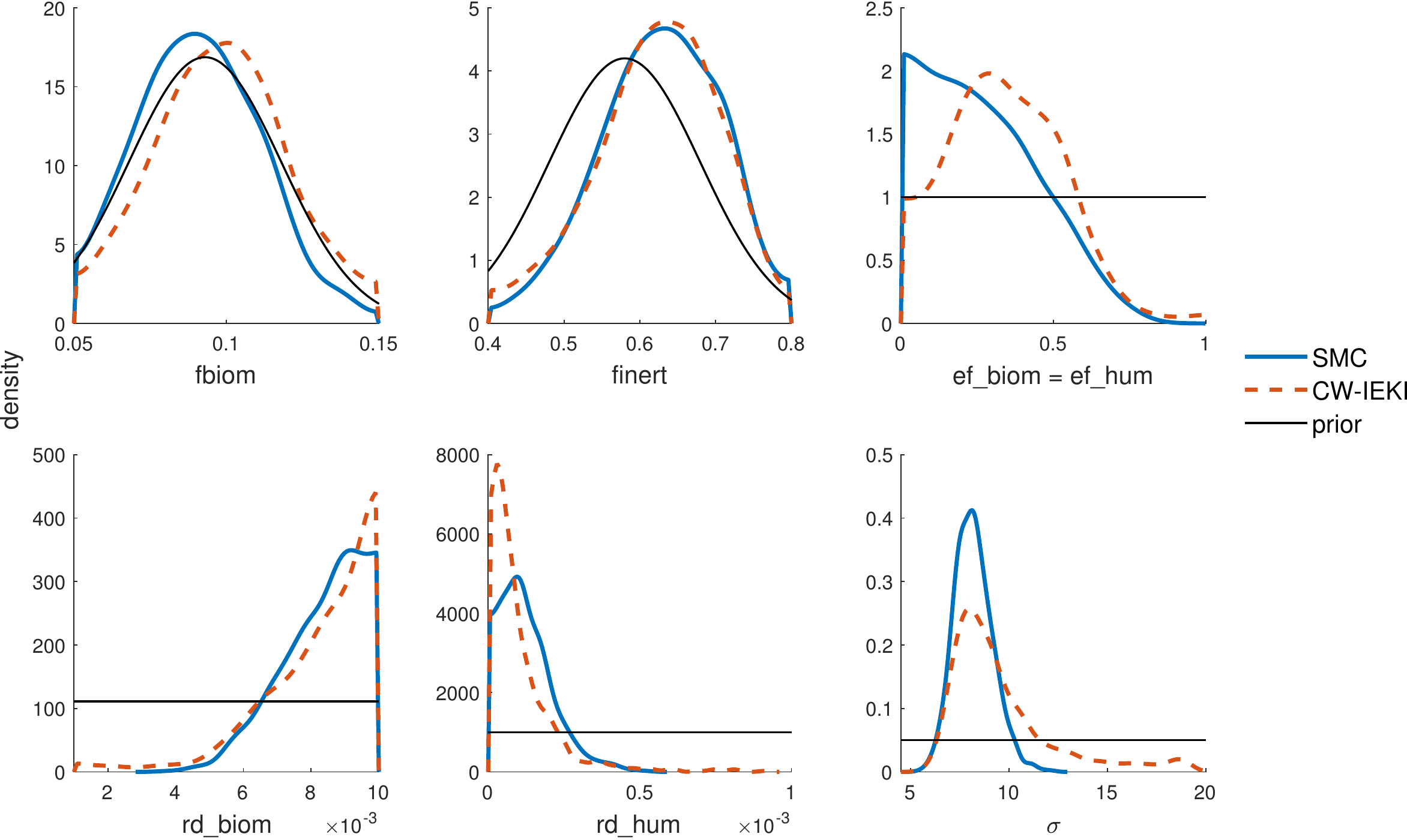}
	\caption{Marginal posterior density plots for the six parameter APSIM model applied to the real data.}
	\label{fig:5Params_densities}
\end{figure}

\begin{figure}[htp]
	\centering
	\includegraphics[scale=0.6]{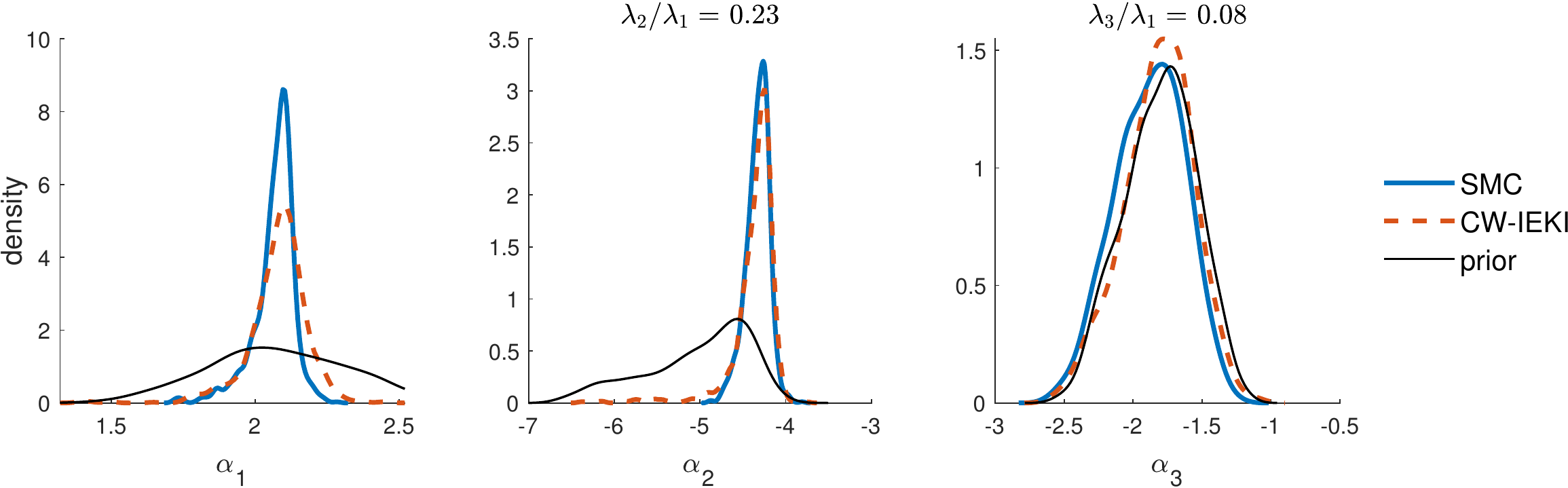}
	\caption{Marginal posterior density plots of the natural logarithm of the three stiffest eigenparameters for the six parameter APSIM model applied to the real data. Note that the uncertainty parameter $\sigma$ is excluded from the analysis of sloppiness, and $\lambda_k$ is the eigenvalue associated with eigenvector $v_k$ in equation \eqref{eq:eigenparameters}. The logarithm of the eigenparameters are calculated based on samples from the prior (black), CW-IEKI (dashed red-orange) and SMC (blue) using a sensitivity matrix calculated using the SMC samples.}
	\label{fig:5Params_densities_eigenparameters}
\end{figure}

\begin{figure}[htp]
	\hspace*{-2cm}    
	\centering
	\includegraphics[scale=0.6]{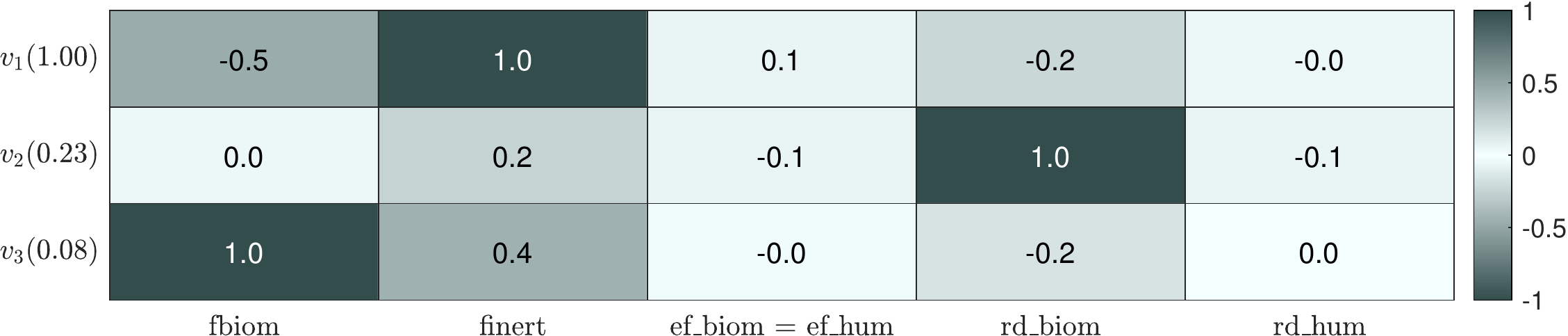} 
	\caption{Eigenvectors of the three stiffest eigenparameters for the six parameter APSIM model applied to the real data. These results are based on the SMC posterior samples. The labels on the left-hand side correspond to $v_k(\lambda_k\slash\lambda_1)$, where $v_k$ and $\lambda_k$ are the eigenvector and associated eigenvalue of eigenparameter $k$, respectively. The shade of the cells in row $k$ indicate the relative contribution ${(v_k)}_j$ of the $j$th parameter to eigenparameter $k$ --- parameters with darker colours have the greatest contribution.}
	\label{fig:5Params_eigenvector}
\end{figure}

\begin{figure}[htp]
	\centering
	\includegraphics[scale=0.6]{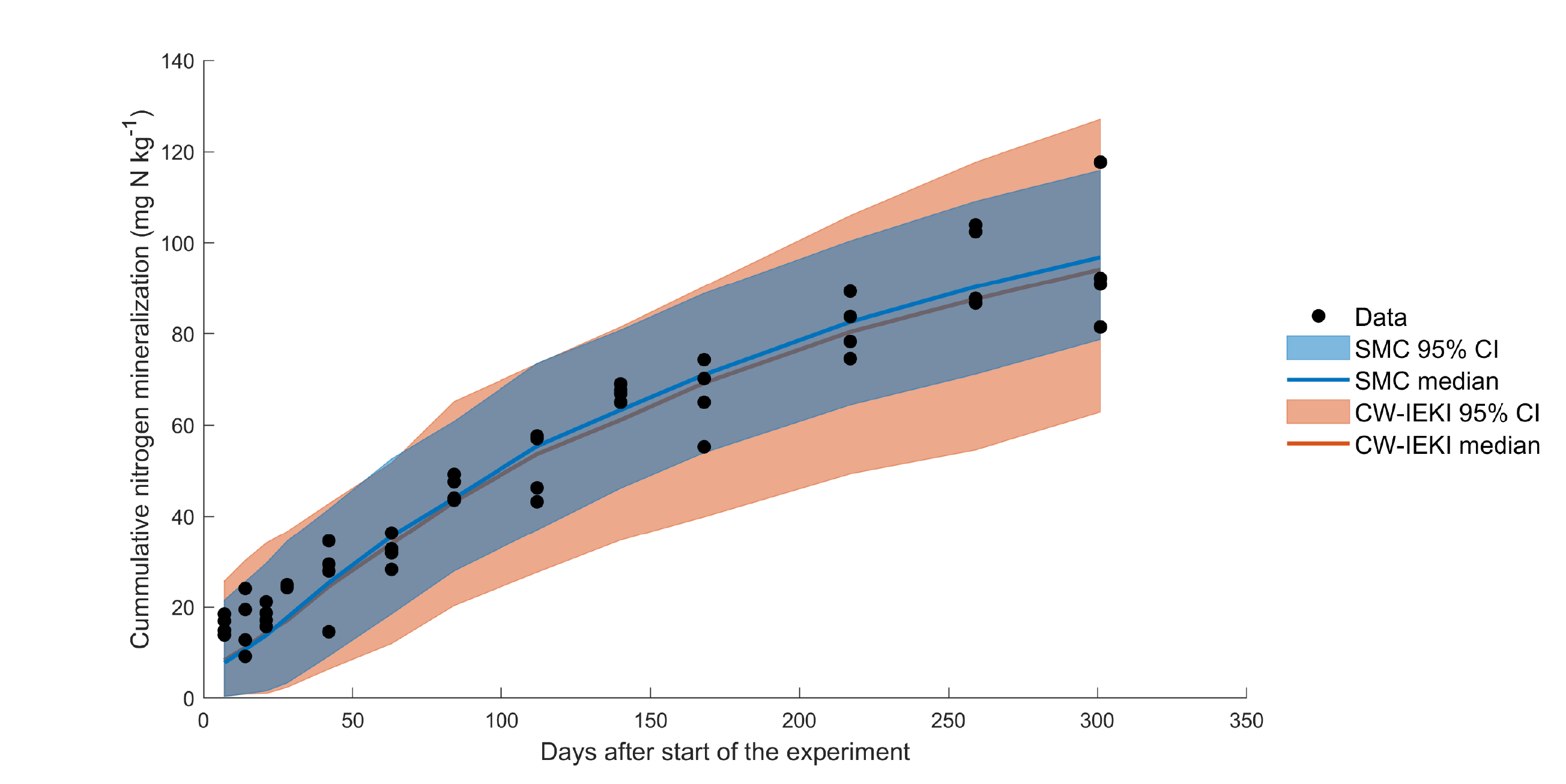}
	\caption{Comparison of the real data to the median and $95\%$ central credible intervals for the posterior predictive distribution of cumulative nitrogen mineralisation obtained from the six parameter APSIM model fitted to this data. Models were fitted using CW-IEKI (red-orange) and likelihood tempering SMC (blue).}
	\label{fig:5Params_postpred}
\end{figure}

Figure \ref{fig:5Params_densities} shows the marginal posterior density plots for the six parameter APSIM model applied to the real data. Figures \ref{fig:5Params_densities_eigenparameters} and \ref{fig:5Params_eigenvector} show the marginal densities and eigenvectors of the three stiffest eigenparameters, and Figure \ref{fig:5Params_postpred} shows the posterior predictive distributions. Unlike the three parameter model, the CW-IEKI and SMC marginal posterior densities have different means for some of the parameters. The predictive performance of CW-IEKI and SMC are relatively similar for this example however, except that the CW-IEKI results have greater uncertainty. This is also shown in the marginal posterior for $\sigma$, where CW-IEKI retains larger values of $\sigma$ in its posterior approximation compared to SMC.

Based on the eigenvectors in Figure \ref{fig:5Params_eigenvector}, the parameters $\textrm{ef\_biom = ef\_hum}$ and rd\_hum do not contribute significantly to the model fit. Interestingly, the CW-IEKI marginal posteriors for these two parameters show the greatest bias compared to the SMC results. Overall, CW-IEKI gives a reasonably good fit for the model. It is also around $33$ times faster than SMC with $9000$ evaluations of $G(\cdot)$ compared to $301000$.

\subsubsection*{Six parameter APSIM model applied to the simulated data}

\begin{figure}[htp]
	\centering
	\includegraphics[scale=0.6]{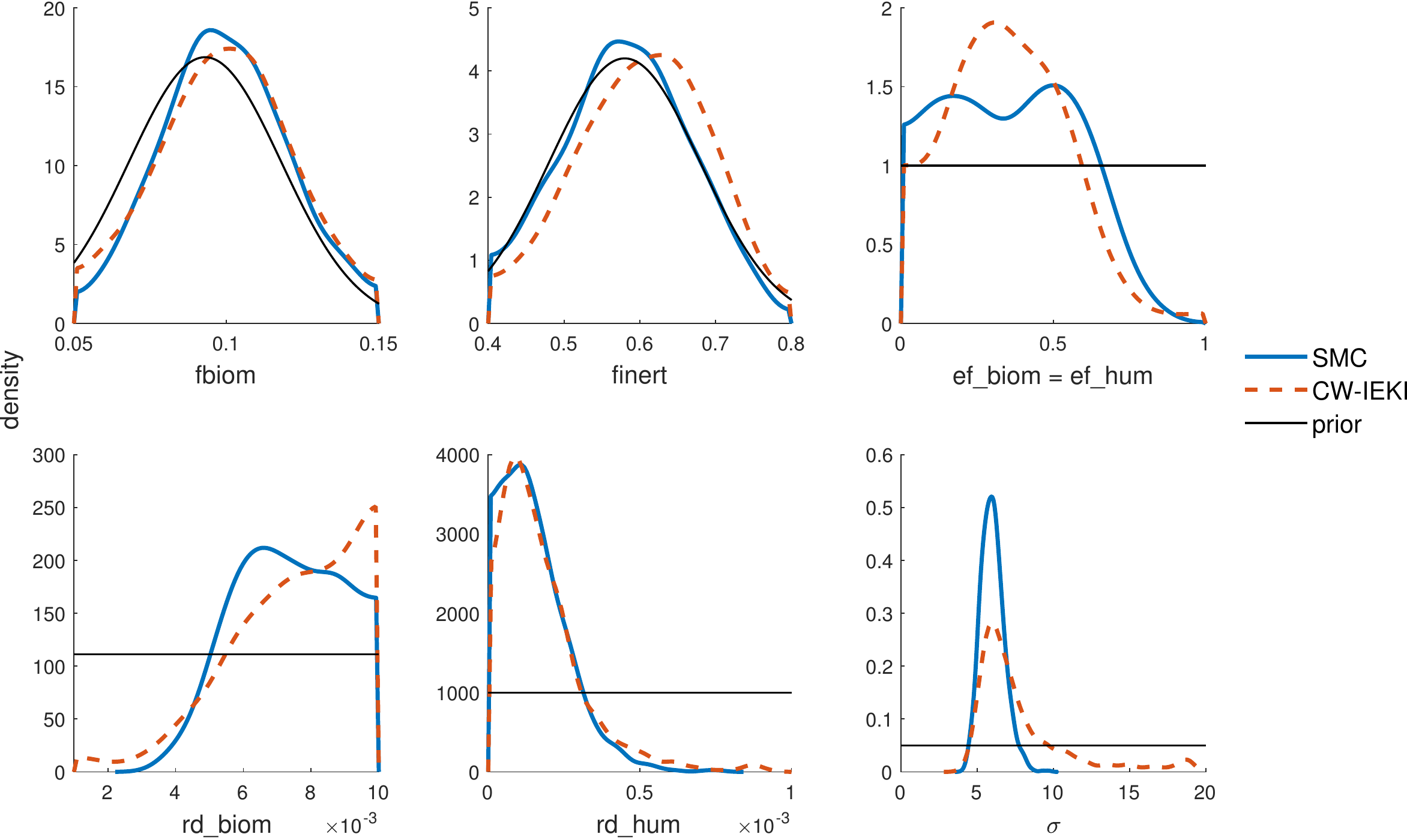}
	\caption{Marginal posterior density plots for the six parameter APSIM model applied to the simulated data.}
	\label{fig:5Params_sim_densities}
\end{figure}

\begin{figure}[htp]
	\centering
	\includegraphics[scale=0.6]{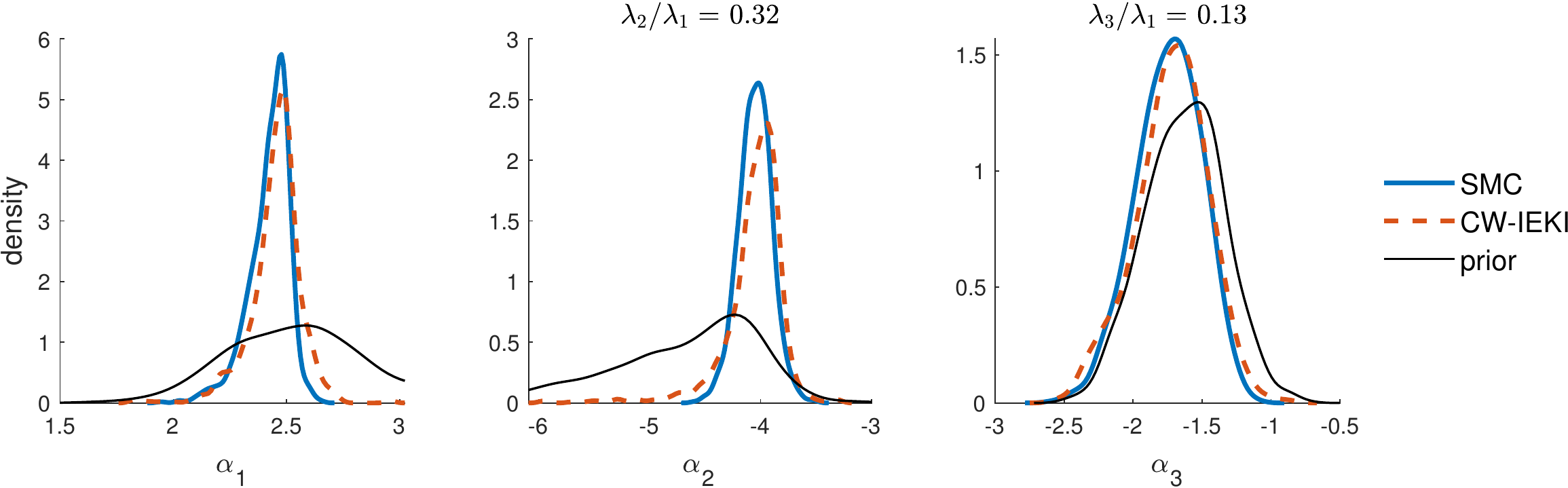}
	\caption{Marginal posterior density plots of the natural logarithm of the three stiffest eigenparameters for the six parameter APSIM model applied to the simulated data. Note that the uncertainty parameter $\sigma$ is excluded from the analysis of sloppiness, and $\lambda_k$ is the eigenvalue associated with eigenvector $v_k$ in equation \eqref{eq:eigenparameters}. The logarithm of the eigenparameters are calculated based on samples from the prior (black), CW-IEKI (dashed red-orange) and SMC (blue) using a sensitivity matrix calculated using the SMC samples.}
	\label{fig:5Params_sim_densities_eigenparameters}
\end{figure}

\begin{figure}[htp]
	\hspace*{-2cm}    
	\centering
	\includegraphics[scale=0.6]{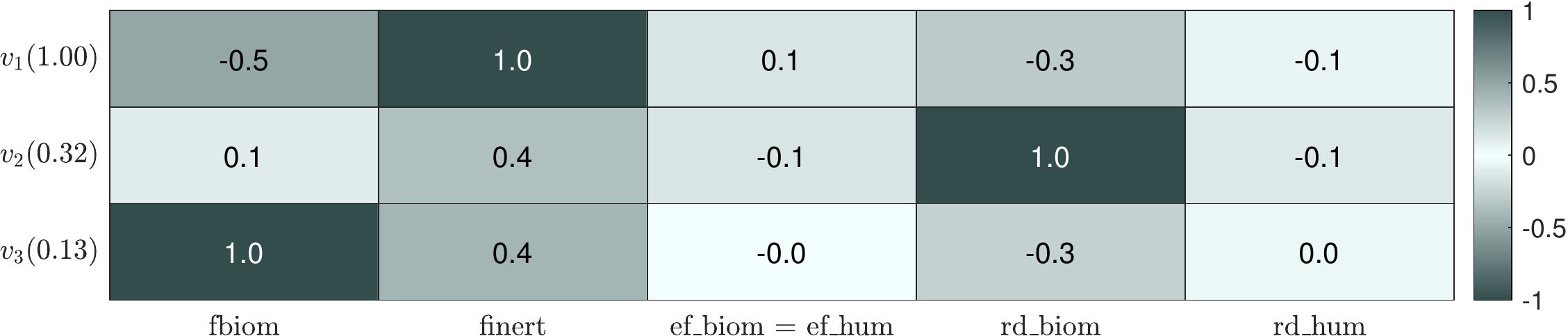} 
	\caption{Eigenvectors of the three stiffest eigenparameters for the six parameter APSIM model applied to the simulated data. These results are based on the SMC posterior samples. The labels on the left-hand side correspond to $v_k(\lambda_k\slash\lambda_1)$, where $v_k$ and $\lambda_k$ are the eigenvector and associated eigenvalue of eigenparameter $k$, respectively. The shade of the cells in row $k$ indicate the relative contribution ${(v_k)}_j$ of the $j$th parameter to eigenparameter $k$ --- parameters with darker colours have the greatest contribution.}
	\label{fig:5Params_sim_eigenvector}
\end{figure}

\begin{figure}[htp]
	\centering
	\includegraphics[scale=0.6]{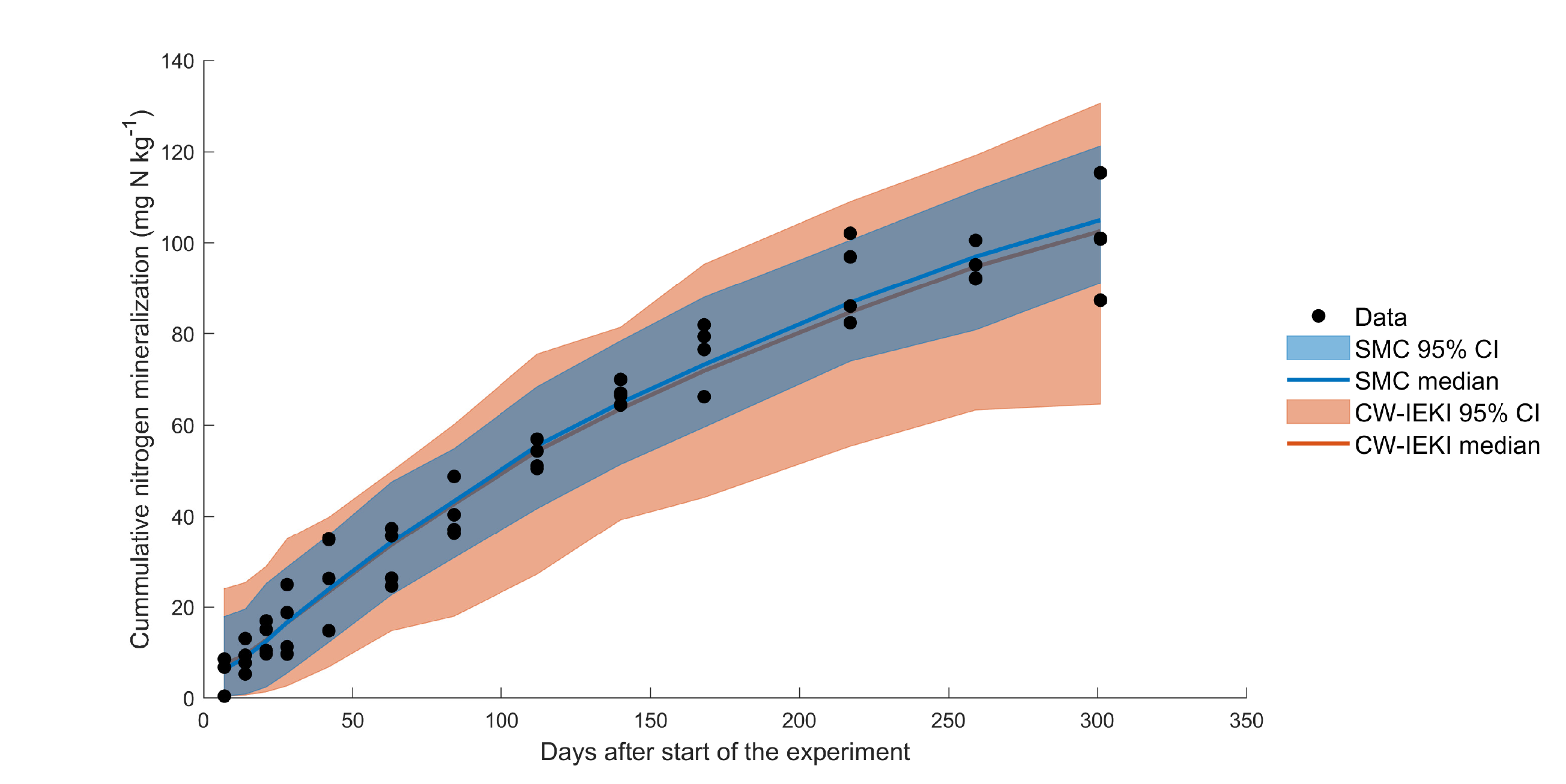}
	\caption{Comparison of the simulated data to the median and $95\%$ central credible intervals for the posterior predictive distribution of cumulative nitrogen mineralisation obtained from the six parameter APSIM model fitted to this data. Models were fitted using CW-IEKI (red-orange) and likelihood tempering SMC (blue).}
	\label{fig:5Params_sim_postpred}
\end{figure}

Figure \ref{fig:5Params_sim_densities} shows the marginal posterior densities of the six parameter APSIM model applied to the simulated data. As before, SMC and CW-IEKI have similar results, except that CW-IEKI has posterior support for larger values of $\sigma$. Thus, this simulation demonstrates that larger support for $\sigma$ from CW-IEKI is not an artefact of model misspecification, as the data used here is simulated from the six parameter APSIM model. Figures \ref{fig:5Params_sim_densities_eigenparameters} and \ref{fig:5Params_sim_eigenvector} show the densities and eigenvectors of the three stiffest eigenparameters respectively. The eigenparameter densities are very similar for SMC and CW-IEKI, indicating that CW-IEKI gives a relatively good fit for the model, and the eigenvectors again show that $\textrm{ef\_biom = ef\_hum}$ and rd\_hum have little influence on the model fit. The posterior predictive distribution in Figure \ref{fig:5Params_sim_postpred} also shows similar performance between SMC and CW-IEKI, except that CW-IEKI has much greater uncertainty. On this example, CW-IEKI is around 37 times faster than SMC with $11000$ evaluations of $G(\cdot)$ compared to $411000$.

\subsection{Model Example 2: Predicting Seagrass Decline}

The second model predicts shoot density decline in seagrass due to cumulative stress from water temperature and light \citep{Adams2020}. The model takes, as input, light, temperature and time period of stress, and outputs photosynthesis rates and changes in shoot density over time. The model has 18 model parameters and 5 noise parameters. Several of these parameters have different values for specific instantaneous temperatures $T$ and mean daily temperatures $\overline{T}$ \citep[see][for full model and parameter details]{Adams2020}. Uniform priors are used for all parameters. See Table 1 for the parameter units and prior bounds.

The model is calibrated to net photosynthesis data \citep{Collier2018} and shoot density data \citep{Collier2016} separately for three species of tropical seagrass from the Great Barrier Reef --– \textit{Cymodocea serrulata}, \textit{Halodule uninervis} and \textit{Zostera muelleri}. In the likelihood function for model-data calibration it is assumed that measurement noise present in net photosynthesis observations at a given temperature $T$ are normally distributed with standard deviation $\sigma_P(T)$. Similarly, measurement noise in shoot density observations is assumed to be normally distributed with standard deviation $\sigma_S$ (albeit with some modifications to account for when observed shoot density declines to zero, see Appendix B of \citealp{Adams2020} for further details).

\begin{table}[htp]
	\centering
	\small
	\begin{tabular}{|c|c|c|c|c|}
		\hline
		Parameter & unit & temperatures ($^{\circ}\textrm{C}$) & lower bound & upper bound(s) \\
		\hline
		$\mu_{\textrm{net,max}}(\overline{T})$ & $\textrm{d}^{-1}$ & $\overline{T} \in \{21.9, 27.9\}$ & $-0.02$ & $0.02$\\
		$C_{\textrm{other loss}}(\overline{T})$ & mg C $\textrm{g}^{-1}$ DW $\textrm{h}^{-1}$ & $\overline{T} \in \{21.9, 27.9\}$ & $0$ & $2.5$\\
		$I_k(T)$ & $\mu$mol m$^{-2}$ s$^{-1}$ & $T\in\{21, 25, 30, 35\}$ & $0$ & $1000$\\
		$R(T)$ & mg C $\textrm{g}^{-1}$ DW $\textrm{h}^{-1}$ & $T\in\{21, 25, 30, 35\}$ & $0$ & $2.5$\\
		$P_{\textrm{max}}(T)$ & mg C $\textrm{g}^{-1}$ DW $\textrm{h}^{-1}$ & $T\in\{21, 25, 30, 35\}$ & $0$ & $20$ (Zm) and $10$ (Cs, Hu)\\
		$\sigma_P(T)$ & mg C $\textrm{g}^{-1}$ DW $\textrm{h}^{-1}$ & $T\in\{21, 25, 30, 35\}$ & $0$ & $2.5$\\
		$k$ & $\textrm{d}^{-1}\slash$ mg C $\textrm{g}^{-1}$ DW $\textrm{h}^{-1}$ & - & $0$ & $0.05$\\
		$S_0$ & shoots\slash pot & - & $0$ & $100$ (Zm), $20$ (Cs) and $50$ (Hu) \\
		$\sigma_S$ & shoots\slash pot & - & $0$ & $20$\\
		\hline
	\end{tabular}
	\caption{Units and prior bounds of all $23$ parameters of the seagrass model. The noise parameters are $\phi = \{\sigma_P(21), \sigma_P(25), \sigma_P(30), \sigma_P(35), \sigma_S\}$, and $\theta$ is comprised of the remaining parameters. In the final column, Cs, Hu and Zm refers to the species \textit{C. serrulata}, \textit{H. uninervis} and \textit{Z. muelleri} respectively.}
	\label{tab:seagrass_parameters}
\end{table}

\begin{figure}[htp]
	\centering
	\includegraphics[scale=0.6]{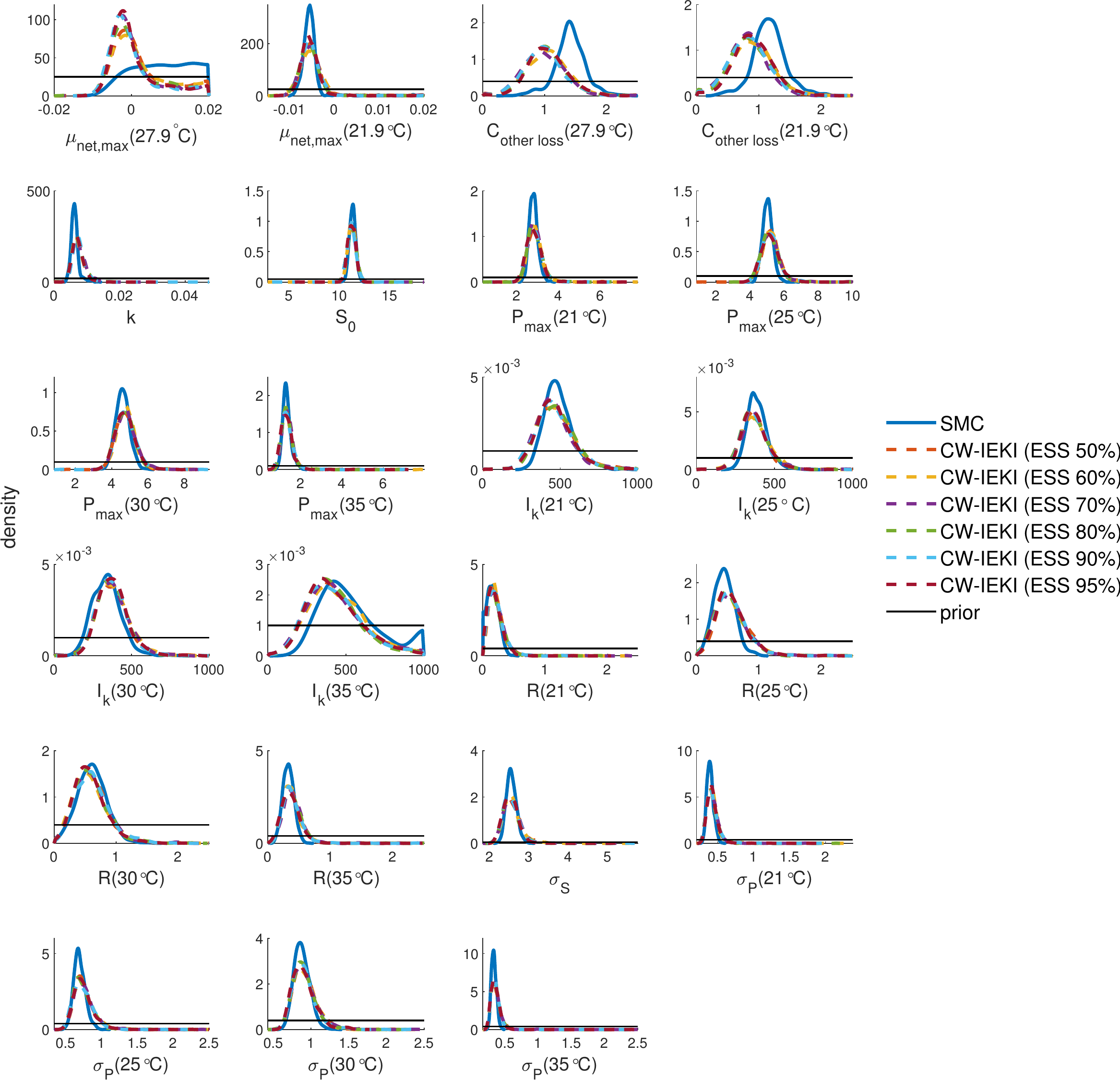}
	\caption{Marginal posterior density plots for the seagrass model applied to the \textit{C. serrulata} data.}
	\label{fig:Cs_densities}
\end{figure}

For brevity, all results shown in this section are for \textit{C. serrulata}. Results for \textit{H. uninervis} and \textit{Z. muelleri} are provided in Appendix \ref{A:seagrass}. On this model, we also test the impact of the target ESS threshold on the accuracy of CW-IEKI. Figure \ref{fig:Cs_densities} shows the marginal posterior densities of the parameters for SMC and CW-IEKI with the different ESS targets. For the majority of the parameters, the SMC and CW-IEKI densities are very similar. The target ESS threshold therefore appears to have little impact on the results.

\begin{figure}[htp]
	\hspace*{-2cm}    
	\centering
	\includegraphics[scale=0.6]{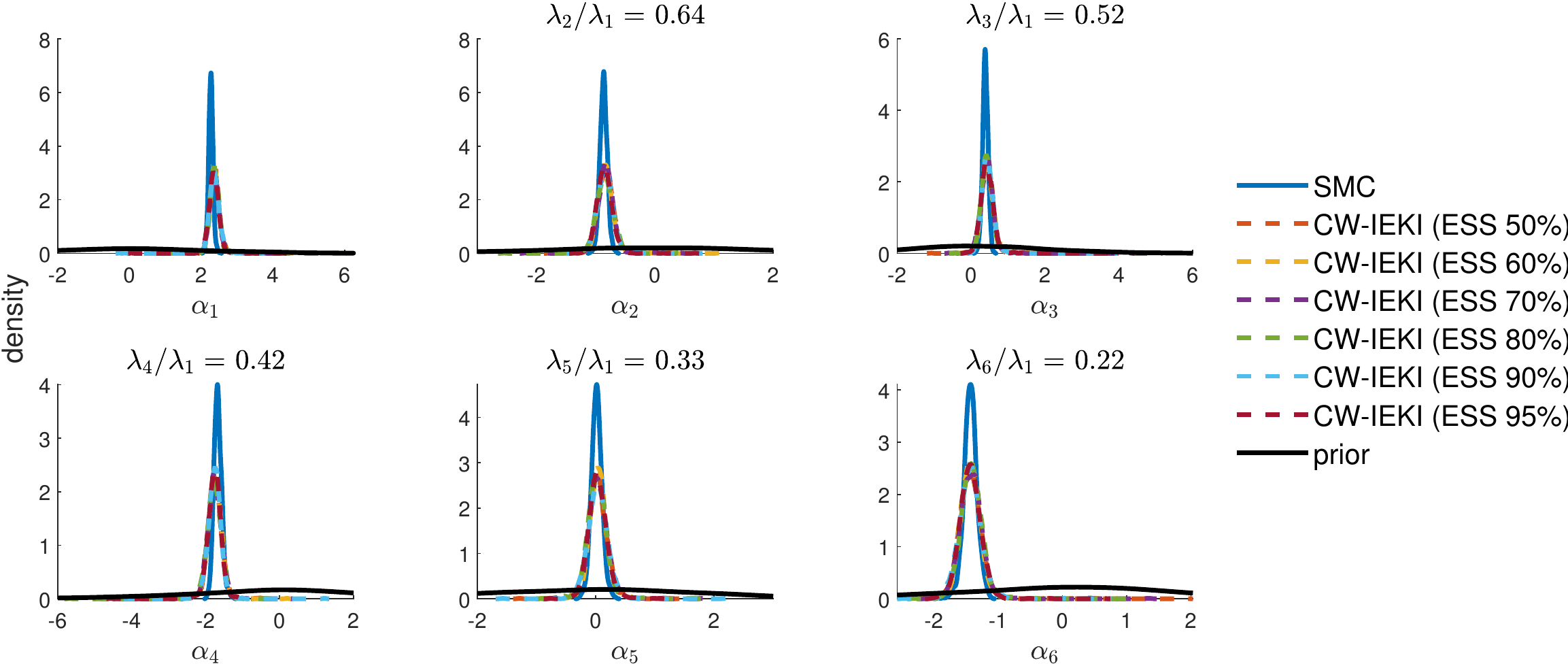} 
	\caption{Marginal density plots of the six stiffest eigenparameters (calculated using equation \eqref{eq:eigenparameters_seagrass}) for the seagrass model applied to the \textit{C. serrulata} data. Note that the uncertainty parameters in $\phi$ are excluded from the analysis of sloppiness, and $\lambda_k$ is the eigenvalue associated with eigenvector $v_k$ in equation \eqref{eq:eigenparameters_seagrass}. The eigenparameters are calculated based on samples from the prior (black), CW-IEKI (dashed) and SMC (blue) using a sensitivity matrix calculated using the SMC samples.}
	\label{fig:Cs_eigenparameter}
\end{figure}

\begin{figure}[htp]
	\hspace*{-2cm}    
	\centering
	\includegraphics[scale=0.6]{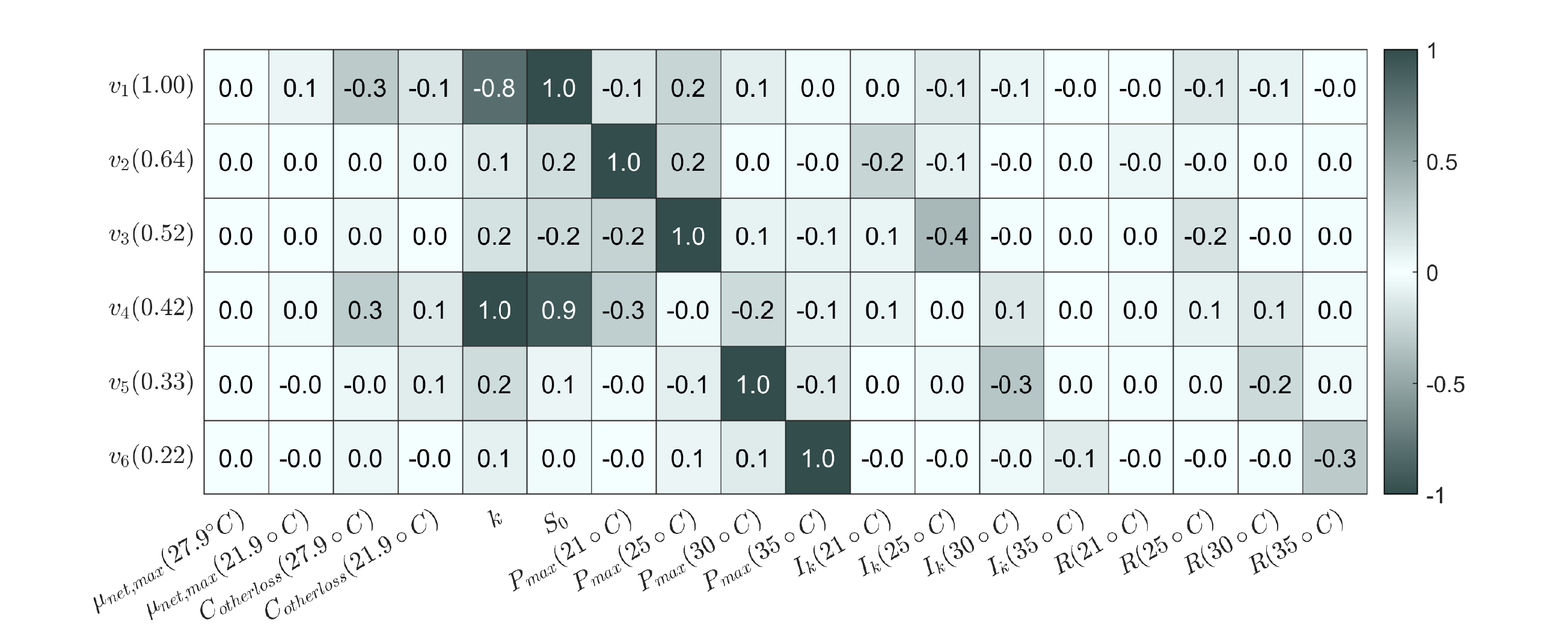} 
	\caption{Eigenvectors of the six stiffest eigenparameters for the seagrass model applied to the \textit{C. serrulata} data. These results are based on the SMC posterior samples. The labels on the left-hand side correspond to $v_k(\lambda_k\slash\lambda_1)$, where $v_k$ and $\lambda_k$ are the eigenvector and associated eigenvalue of eigenparameter $k$, respectively. The shade of the cells in row $k$ indicate the relative contribution ${(v_k)}_j$ of the $j$th parameter to eigenparameter $k$ --- parameters with darker colours have the greatest contribution.}
	\label{fig:Cs_eigenvectors}
\end{figure}

As the parameters $\mu_{\textrm{net,max}}(21.9)$ and $\mu_{\textrm{net,max}}(27.9)$ are bounded between $-0.02$ and $0.02$ (see Table \ref{tab:seagrass_parameters}), the log-transform cannot be used when performing the analysis of model sloppiness. Instead, we rescale all the model parameter to be between $[0, 1]$ using the prior bounds, and then apply a logit transformation to map these values back to $[-\infty, \infty]$. The sensitivity matrix is given by the inverse of the covariance of the logit-transformed posterior samples from SMC, and the eigenparameters are given by
\begin{align}
	\alpha_k = \sum_{j=1}^{d_{\theta}}{(v_k)_j\log{\left(\frac{\dot{\theta}_j}{1 - \dot{\theta}_j}\right)}}, \quad \dot{\theta}_j = \frac{(\theta_{j} - a_j)}{(b_j - a_j)},
	\label{eq:eigenparameters_seagrass}
\end{align}
where $(v_k)_j$ is the $j$th element of the $k$th normalised eigenvector, $\theta_j$ is the $j$th parameter, $a_j$ is the prior lower bound of parameter $j$ and $b_j$ is the prior upper bound of parameter $j$. Figures \ref{fig:Cs_eigenparameter} and \ref{fig:Cs_eigenvectors} show the marginal densities and eigenvectors of the six stiffest eigenparameters. The densities of these eigenparameters are similar for SMC and CW-IEKI, although again, the CW-IEKI results have greater uncertainty. Based on the eigenvectors in Figure \ref{fig:Cs_eigenvectors}, the parameters $\mu_{\textrm{net,max}}(21.9)$, $\mu_{\textrm{net,max}}(27.9)$, $C_{\textrm{other loss}}(21.9)$ and $C_{\textrm{other loss}}(27.9)$ do not significantly influence the model fit. As with the six parameter APSIM model, the CW-IEKI marginal posteriors for less influential parameters show the greatest bias.

\begin{figure}[htp]
	\hspace*{-2cm}    
	\centering
	\includegraphics[scale=0.6]{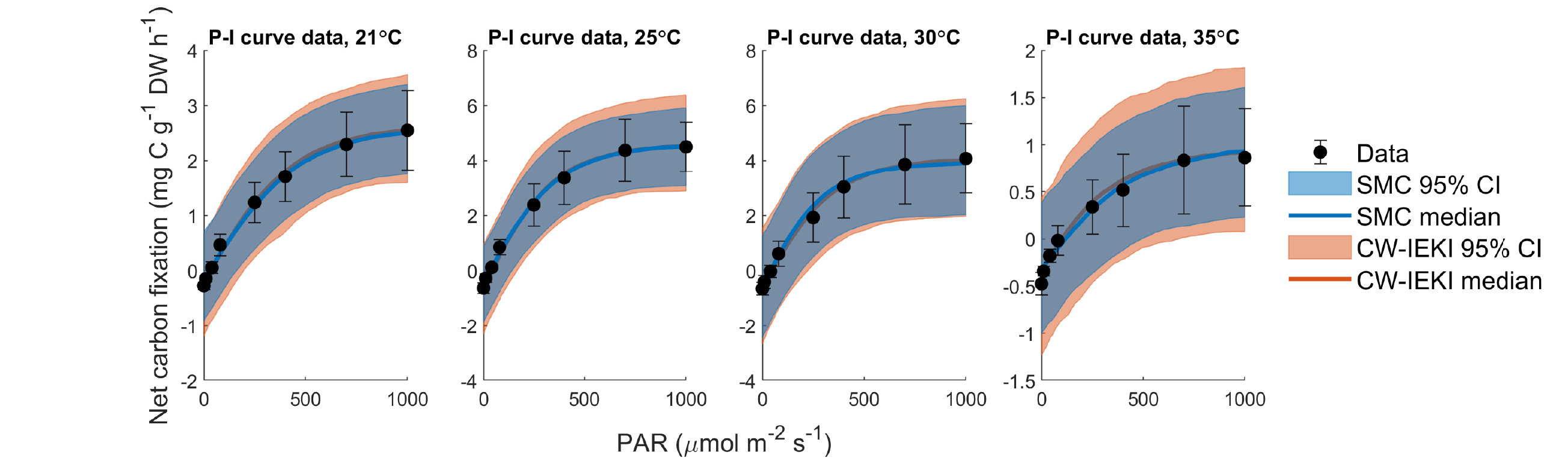}
	\caption{Comparison of the \textit{C. serrulata} data to the median and $95\%$ central credible intervals for the posterior predictive distribution of net carbon fixation obtained from the seagrass model fitted to this data. Models were fitted using CW-IEKI (red-orange) and likelihood tempering SMC (blue). P-I $=$ photosynthesis-irradiance and PAR $=$ photosynthetically active radiation \citep[see][]{Adams2020}.}
	\label{fig:Cs_postpred_PI}
\end{figure}

\begin{figure}[htp]
	\hspace*{-2cm}    
	\centering
	\includegraphics[scale=0.6]{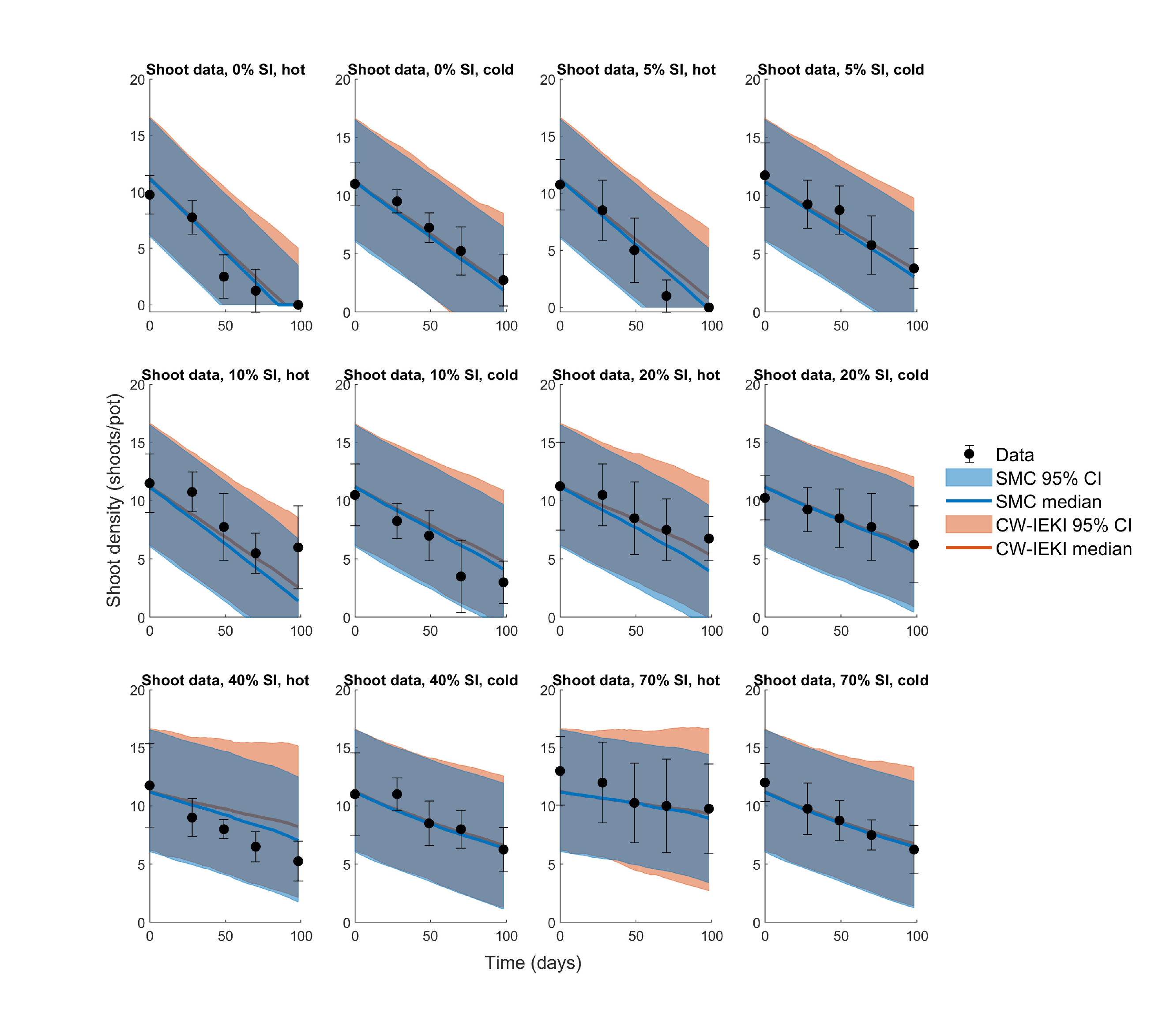}
	\caption{Comparison of the \textit{C. serrulata} data to the median and $95\%$ central credible intervals for the posterior predictive distribution of shoot density obtained from the seagrass model fitted to this data. Models were fitted using CW-IEKI (red-orange) and likelihood tempering SMC (blue). SI $=$ surface irradiance and ``hot" or ``cold" indicates the temperature conditions under which the seagrass data was collected \citep{Adams2020}.}
	\label{fig:Cs_postpred_SD}
\end{figure}

Figure \ref{fig:Cs_postpred_PI} shows the posterior predictive plots of the net carbon fixation using SMC and CW-IEKI with an ESS target of $50\%$, and Figure \ref{fig:Cs_postpred_SD} shows the posterior predictive plots of the shoot density decline. The predictive performance of SMC and CW-IEKI are fairly similar for this example, except that the CW-IEKI predictions have greater uncertainty. As with the density plots, there is little difference between the posterior predictive plots for an ESS target threshold of 50\% and higher ESS targets (not shown). Table \ref{tab:seagrass_computation} shows the computation cost for SMC and CW-IEKI. For an ESS target of $50\%$, CW-IEKI is approximately $40$ times faster than SMC.

The results suggest that CW-IEKI gives a good fit for predicting shoot density decline and carbon fixation for \textit{C. serrulata}. CW-IEKI also gives a good fit for \textit{H. uninervis}, but not for \textit{Z. muelleri} (see Appendix \ref{A:seagrass}). The relatively poor fit for the latter may be a result of the likelihood not being strictly Gaussian due to the modifications that ensure the predicted shoot density remains greater than or equal to $0$. As a result of these modifications, the likelihood function is close to Gaussian for higher shoot density values, but deviates strongly when the shoot density declines to $0$, which is more often the case for \textit{Z. muelleri} than for the other seagrass species.

\begin{table}[htp]
	\centering
	\begin{tabular}{|c|c|cccccc|}
		\hline
		Method & SMC & \multicolumn{6}{c|}{CW-IEKI} \\
		ESS target threshold & $50\%$ & $95\%$ & $90\%$ & $80\%$ & $70\%$ & $60\%$ & $50\%$ \\
		\hline
		$G(\cdot)$ evaluations & 799000 & 87000 & 60000 & 39000 & 30000 & 24000 & 20000 \\
		Approximate speed-up & 1.00 & 9.18 & 13.32 & 20.49 & 26.63 & 33.29 & 39.95 \\
		\hline 
	\end{tabular}
	\caption{Total and relative number of evaluations of $G(\cdot)$ for SMC and CW-IEKI with different ESS target thresholds. Results are for the seagrass model applied to the \textit{C. serrulata} data. Note that the total number of evaluations of $G(\cdot)$ is a multiple of the number of samples $N = 1000$.}
	\label{tab:seagrass_computation}
\end{table}

\subsection{Model Example 3: Predicting Coral Calcification Rates}

The final model predicts coral calcification rates by simulating the transport and reaction of relevant chemical species and metabolic fluxes from seawater to the coral skeleton. It is assumed that there are two layers between the seawater and the coral skeleton: the coelenteron and the extracellular calcifying medium (ECM).

The main reactions considered are photosynthesis and respiration (seawater $\leftrightarrow$ coelenteron), passive transport processes (seawater $\leftrightarrow$ coelenteron $\leftrightarrow$ ECM), membrane transport processes (coelenteron $\leftrightarrow$ ECM) and aragonite precipitation and dissolution (ECM $\leftrightarrow$ coral skeleton). The two membrane transport pumps modelled as part of the membrane transport processes are a Ca-ATPase pump and a bicarbonate anion transport (BAT) pump. 

The reactions are modelled by a system of ordinary differential equations (ODEs), and measurement error is assumed to be Gaussian with standard deviation $\sigma$. The calcification rate predictions of the model are obtained from the steady state solution of the ODEs --- these are compared to the data for calibration. There are a total of $21$ unknown parameters which correspond to the passive transport processes, the membrane transport processes and the measurement error variance. Uniform priors are used for all parameters. Table \ref{tab:coral_parameters} shows the parameter units and prior bounds. See \citet{Galli2018} for more detail about the model and the values of the remaining parameters, and \citet{Vollert2022} for an application of SMC and analysis of model sloppiness to this model-data calibration problem.

The model is applied to data from \citet{RodolfoMetalpa2010} measuring the photosynthesis, respiration and calcification of the Mediterranean coral \textit{C. caespitosa}. The data was measured at winter and summer baseline ($13.4$ and $21.7^{\circ}\textrm{C}$) and elevated ($16.4$ and $24.5^{\circ}\textrm{C}$) temperatures, two different pCO$_{2}$ levels ($400$ and $700$ ppm), and under light and dark conditions, giving $16$ data points overall.

\begin{figure}[H]
	\centering
	\includegraphics[scale=0.6]{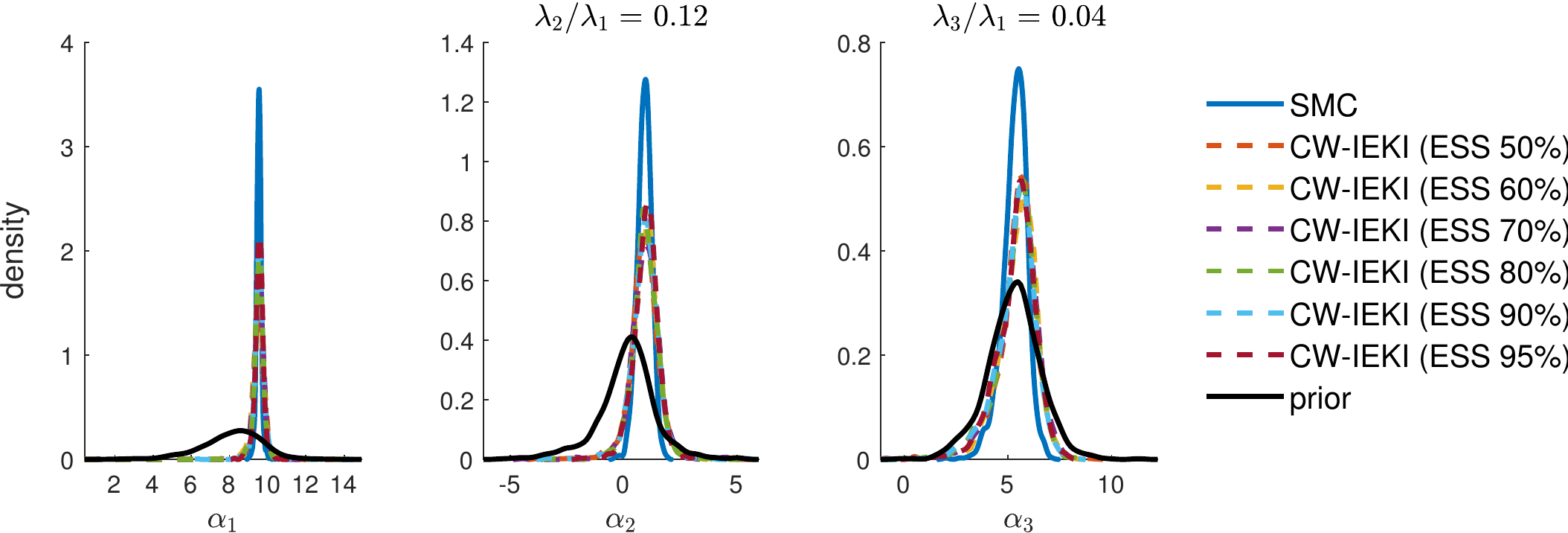}
	\caption{Marginal posterior density plots of the natural logarithm of the three stiffest eigenparameters for the coral model. CW-IEKI results use different ESS target thresholds. Note that the uncertainty parameter $\sigma$ is excluded from the analysis of sloppiness, and $\lambda_k$ is the eigenvalue associated with eigenvector $v_k$ in equation \eqref{eq:eigenparameters}. The logarithm of the eigenparameters are calculated based on samples from the prior (black), CW-IEKI (dashed) and SMC (blue) using a sensitivity matrix calculated using the SMC samples.}
	\label{fig:coral_sloppy_densities}
\end{figure}

\begin{figure}[H]
	\centering
	\includegraphics[scale=0.6]{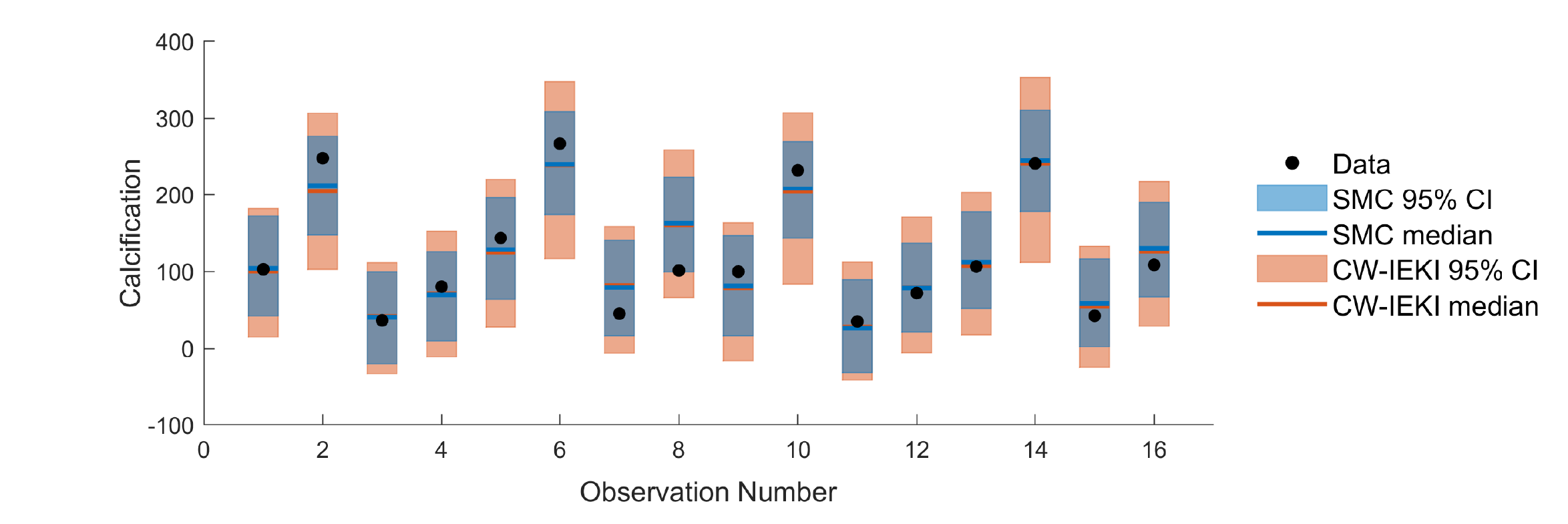}
	\caption{Comparison of the coral data to the median and $95\%$ central credible intervals for the posterior predictive distribution of coral calcification obtained from the coral model fitted to this data. Models were fitted using CW-IEKI (red-orange) and likelihood tempering SMC (blue).}
	\label{fig:coral_postpred}
\end{figure}

Figure \ref{fig:coral_sloppy_densities} shows the marginal densities of the three stiffest eigenparameters, Figure \ref{fig:coral_postpred} shows the posterior predictive distribution and Table \ref{tab:coral_computation} shows the computation cost for SMC and CW-IEKI. Due to the limited data available for this model, the marginal posterior densities for SMC and CW-IEKI are close to the prior (see Appendix \ref{A:coral}). In contrast, the marginal densities of the stiffest eigenparameters shown in Figure \ref{fig:coral_sloppy_densities} are much more informative. As with previous examples, the predictive performance of SMC and CW-IEKI are similar, except that the CW-IEKI results have greater uncertainty (Figure \ref{fig:coral_postpred}). Again, changing the ESS target threshold for CW-IEKI makes very little difference to the results. For a target threshold of $50\%$, CW-IEKI is almost $24$ times faster than SMC (Table \ref{tab:coral_computation}).

\begin{table}[H]
	\centering
	\small
	\begin{tabular}{|c|c|c|c|c|}
		\hline
		Reaction & Parameter & Unit & Lower bound & Upper bound \\
		\hline
		\multirow{3}{2cm}{Passive transport processes} & $k_{{CO}_2}$ & cm s$^{-1}$ & $0$ & $0.1$ \\
		& $k_{pp}$ & cm s$^{-1}$ & $0$ & $0.1$ \\
		& $s$ & cm s$^{-1}$ & $0$ & $0.1$ \\
		\hline
		\multirow{10}{2cm}{Ca-ATPase mechanism} & $\alpha$ & - & $0$ & $1$ \\
		& $\beta$ & - & $0$ & $1$ \\
		& $v_{H_c}$ & cm s$^{-1}$ & $0$ & $250$ \\
		& $E_{0_c}$ & $\mu$mol cm$^{-2}$ & $0$ & $1.2\times 10^7$ \\
		& $k_{1f_c}$ & cm$^4$ s $\mu$mol$^{-2}$ & $0$ & $1.4\times 10^{-4}$ \\
		& $k_{2f_c}$ & s$^{-1}$ & $0$ & $0.5$ \\
		& $k_{3f_c}$ & s$^{-1}$ & $0$ & $800$ \\
		& $k_{1b_c}$ & cm$^2$ $\mu$mol$^{-1}$ & $0$ & $8$ \\
		& $k_{2b_c}$ & s$^{-1}$ & $0$ & $500$ \\
		& $k_{3b_c}$ & cm$^4$ s $\mu$mol$^{-2}$ & $0$ & $1\times 10^{-7}$ \\
		\hline
		\multirow{7}{2cm}{BAT mechanism} & $E_{0_b}$ & $\mu$mol cm$^{-2}$ & $0$ & $1500$ \\
		& $k_{1f_b}$ & cm$^3$ $\mu$mol$^{-1}$ s$^{-1}$ & $0$ & $5 \times 10^{-5}$ \\
		& $k_{2f_b}$ & s$^{-1}$ & $0$ & $0.01$ \\
		& $k_{3f_b}$ & s$^{-1}$ & $0$ & $0.01$ \\
		& $k_{1b_b}$ & s$^{-1}$ & $0$ & $2 \times 10^{-4}$ \\
		& $k_{2b_b}$ & s$^{-1}$ & $0$ & $1 \times 10^{-3}$ \\
		& $k_{3b_b}$ & cm$^3$ $\mu$mol$^{-1}$ s$^{-1}$ & $0$ & $3.5 \times 10^{-9}$ \\
		\hline
		& $\sigma$ & $\mu$mol cm$^{-2}$ h$^{-1}$ & $0$ & $50$ \\
		\hline
	\end{tabular}
	\caption{Units and prior bounds of all $21$ parameters of the coral calcification model. For this model, the uncertainty parameter is $\phi = \sigma$.}
	\label{tab:coral_parameters}
\end{table}

\begin{table}[htp]
	\centering
	\begin{tabular}{|c|c|cccccc|}
		\hline
		Method & SMC & \multicolumn{6}{c|}{CW-IEKI} \\
		ESS target threshold & $50\%$ & $95\%$ & $90\%$ & $80\%$ & $70\%$ & $60\%$ & $50\%$ \\
		\hline
		$G(\cdot)$ evaluations & 166000 & 32000 & 21000 & 13000 & 10000 & 8000 & 7000 \\
		Approximate speed-up & 1.00 & 5.19 & 7.90 & 12.77 & 16.60 & 20.75 & 23.71 \\
		\hline 
	\end{tabular}
	\caption{Total and relative number of evaluations of $G(\cdot)$ for SMC and CW-IEKI with different ESS target thresholds for the coral model. Note that the total number of evaluations of $G(\cdot)$ is a multiple of the number of samples $N = 1000$.}
	\label{tab:coral_computation}
\end{table}

	\section{Discussion} \label{sec:discussion}

	In this paper, we have introduced and tested a new method (CW-IEKI) which extends the IEKI method of \citet{Iglesias2018} to the case where the covariance matrix has unknown elements $\phi$. Our component-wise IEKI approach is completely analogous to likelihood tempering SMC, and is a useful alternative to both MCMC and SMC for static Bayesian models of the form $\mathcal{N}(G(\theta), \Gamma(\phi))$, where $\theta$ and $\phi$ are unknown, and when $G(\theta)$ is expensive to compute. Note that CW-IEKI can also be applied when the covariance matrix is a function of both $\theta$ and $\phi$, as is the case for the six parameter APSIM model applied to the simulated data in Section \ref{sec:apsim}. That is, CW-IEKI can be applied to models of the form $\mathcal{N}(G(\theta), \Gamma(\theta, \phi))$, where $\theta\mid\phi$ is updated using EKI and $\phi\mid\theta$ is updated using MCMC. Even though the inference from CW-IEKI is only unbiased for models with a linear Gaussian likelihood and Gaussian prior, we find in practice that it provides reasonable inference even if the model's likelihood is non-linear Gaussian and its prior is non-Gaussian. Additionally, CW-IEKI generally requires much fewer evaluations of $G(\cdot)$ than MCMC or SMC. 
	
	We compared our method to SMC on three ecological models, all of which have a non-linear Gaussian likelihood and a non-Gaussian prior. The accuracy, predictive performance and computation time, the latter of which is measured by the number of evaluations of the function $G(\cdot)$, were used to assess the performance of our method relative to the unbiased solution from SMC. In the three parameter APSIM model and the coral model, the accuracy of CW-IEKI and SMC were similar, but for the remaining models there was clear bias in the marginal posteriors for some of the parameters. Based on the stiffest eigenparameters however, the model parameters showing the most bias also had little impact on the model fit. Across all models, CW-IEKI had relatively similar predictive performance to SMC --- except that the uncertainty of the predictions was consistently overestimated --- but advantageously required $11$-$40$ times less evaluations of $G(\cdot)$. We also found that increasing the ESS target threshold for CW-IEKI made little difference to its accuracy and predictive performance. 
	
	In all of the examples we found that the point predictions from our novel CW-IEKI method was quite accurate, but the uncertainty intervals were inflated relative to SMC, especially when the number of parameters was increased. Therefore if highly accurate uncertainty quantification or parameter inferences are needed for a given application, then SMC or MCMC may be worth the wait if they are computationally feasible. If exact inferences are desired, CW-IEKI proposals could potentially be used to speed up exact SMC, for example by incorporating them in the delayed-acceptance SMC algorithm of \citet{Bon2021}. The inferences from CW-IEKI could also potentially be improved by following the approach of \citet{Lan2022} to build an emulator $\widehat{G}(\cdot)$ of $G(\cdot)$ using all evaluations of $G(\cdot)$ from CW-IEKI. An MCMC or SMC algorithm can then be used to target the approximate posterior distribution based on this emulator, i.e.\ $\mathcal{N}(y\mid\widehat{G}(\theta), \Gamma(\phi))p(\theta, \phi)$.
		
	An area of future work is to improve the updates in CW-IEKI for the noise parameters $\phi$. Currently, a fixed number of random-walk MCMC iterations are used. Adapting the number of MCMC iterations and using more efficient updates for $\phi$, such as the Metropolis-adjusted Langevin algorithm \citep{Girolami2011} or Hamiltonian Monte Carlo \citep{Betancourt2017}, may improve the performance of the method, especially if $\phi$ is high-dimensional or its elements are highly correlated. Another extension is to apply the CW-IEKI method to the hierarchical setting explored in \citet{Chada2018}. 
	
	Another avenue of future work is to investigate how our approach can be incorporated into the IEKI method of \citet{Duffield2021} for general likelihoods. In this case, it may be possible to update some of the model parameters with IEKI and some with MCMC, depending on the form of the likelihood function. The potential advantage of such a hybrid approach is that it may efficiently improve the accuracy of the final samples, given that the MCMC update targets the exact conditional posterior, while the IEKI portion targets some approximation to the conditional posterior. 
	
	It would also be interesting to incorporate the CW-IEKI method into the data annealing SMC algorithm of \citet{Wu2022}, which currently requires the covariance of the likelihood function to be known. In particular, one extension here is developing a likelihood tempering SMC algorithm with our CW-IEKI method as the forward kernel. It may also be possible to extend their SMC algorithm to general likelihood models, such that a subset of the parameters are updated using the method of \citet{Duffield2021}, and the rest are updated using an MCMC forward kernel. 
	
	\section{Acknowledgments} \label{sec:ack}
	  We thank Maria P. Vilas and Kirsten Verburg for helpful discussions, suggestions and clarifications on the APSIM model used in this paper. We thank Diane Allen, Tom Orton and Phil Bloesch from the Department of Environment and Science for sharing the measured mineralisation data. We thank Catherine Collier for sharing the seagrass mesocosm data. We gratefully acknowledge the computational resources provided by QUT's High Performance Computing and Research Support Group (HPC). Imke Botha was supported by an Australian Research Training Program Stipend and a QUT Centre for Data Science Top-Up Scholarship. Matthew P. Adams was supported by an Australian Research Council Discovery Early Career Researcher Award (DE200100683). Christopher Drovandi was supported by an Australian Research Council Future Fellowship (FT210100260). 

\bibliographystyle{apalike} 
\bibliography{refs}

\begin{thebibliography}{}

\bibitem[Adams et~al., 2020]{Adams2020}
Adams, M.~P., Koh, E. J.~Y., Vilas, M.~P., Collier, C.~J., Lambert, V.~M.,
  Sisson, S.~A., Quiroz, M., McDonald-Madden, E., McKenzie, L.~J., and
  O{\textquotesingle}Brien, K.~R. (2020).
\newblock Predicting seagrass decline due to cumulative stressors.
\newblock {\em Environmental Modelling {\&} Software}, 130:104717.

\bibitem[Allen et~al., 2019]{Allen2019}
Allen, D.~E., Bloesch, P.~M., Orton, T.~G., Schroeder, B.~L., Skocaj, D.~M.,
  Wang, W., Masters, B., and Moody, P.~M. (2019).
\newblock {Nitrogen mineralisation in sugarcane soils in Queensland, Australia:
  I. evaluation of soil tests for predicting nitrogen mineralisation}.
\newblock {\em Soil Research}, 57(7):738.

\bibitem[APHA and AWWA, 2012]{APHA2012}
APHA and AWWA (2012).
\newblock {\em {Standard Methods for the Examination of Water and Wastewater}}.
\newblock American Water Works Association, 22nd edition.

\bibitem[Betancourt, 2017]{Betancourt2017}
Betancourt, M. (2017).
\newblock {A Conceptual Introduction to Hamiltonian Monte Carlo}.
\newblock {\em arXiv preprints}, page arXiv:1701.02434.

\bibitem[Bon et~al., 2021]{Bon2021}
Bon, J.~J., Lee, A., and Drovandi, C. (2021).
\newblock {Accelerating sequential Monte Carlo with surrogate likelihoods}.
\newblock {\em Statistics and Computing}, 31(5).

\bibitem[Burgers et~al., 1998]{Burgers1998}
Burgers, G., van Leeuwen, P.~J., and Evensen, G. (1998).
\newblock {Analysis Scheme in the Ensemble Kalman Filter}.
\newblock {\em Monthly Weather Review}, 126(6):1719--1724.

\bibitem[Chada et~al., 2018]{Chada2018}
Chada, N.~K., Iglesias, M.~A., Roininen, L., and Stuart, A.~M. (2018).
\newblock Parameterizations for ensemble {K}alman inversion.
\newblock {\em Inverse Problems}, 34(5):055009.

\bibitem[Chada et~al., 2020]{Chada2020}
Chada, N.~K., Stuart, A.~M., and Tong, X.~T. (2020).
\newblock {Tikhonov Regularization within Ensemble Kalman Inversion}.
\newblock {\em {SIAM} Journal on Numerical Analysis}, 58(2):1263--1294.

\bibitem[Collier et~al., 2016]{Collier2016}
Collier, C.~J., Adams, M.~P., Langlois, L., Waycott, M., O'Brien, K.~R.,
  Maxwell, P.~S., and McKenzie, L. (2016).
\newblock Thresholds for morphological response to light reduction for four
  tropical seagrass species.
\newblock {\em Ecological Indicators}, 67:358--366.

\bibitem[Collier et~al., 2018]{Collier2018}
Collier, C.~J., Langlois, L., Ow, Y., Johansson, C., Giammusso, M., Adams,
  M.~P., O{\textquotesingle}Brien, K.~R., and Uthicke, S. (2018).
\newblock Losing a winner: thermal stress and local pressures outweigh the
  positive effects of ocean acidification for tropical seagrasses.
\newblock {\em New Phytologist}, 219(3):1005--1017.

\bibitem[{Del Moral} et~al., 2006]{DelMoral2006}
{Del Moral}, P., Doucet, A., and Jasra, A. (2006).
\newblock {Sequential Monte Carlo samplers}.
\newblock {\em Journal of the Royal Statistical Society: Series B (Statistical
  Methodology)}, 68(3):411--436.

\bibitem[Ding and Li, 2021]{Ding2021}
Ding, Z. and Li, Q. (2021).
\newblock {Ensemble Kalman Sampler: Mean-field Limit and Convergence Analysis}.
\newblock {\em {SIAM} Journal on Mathematical Analysis}, 53(2):1546--1578.

\bibitem[Duffield and Singh, 2021]{Duffield2021}
Duffield, S. and Singh, S.~S. (2021).
\newblock {Ensemble Kalman Inversion for General Likelihoods}.
\newblock {\em arXiv preprints}, page arXiv:2110.03034.

\bibitem[Evensen, 1994a]{Evensen1994}
Evensen, G. (1994a).
\newblock Sequential data assimilation with a nonlinear quasi-geostrophic model
  using {Monte Carlo} methods to forecast error statistics.
\newblock {\em Journal of Geophysical Research}, 99(C5):10143.

\bibitem[Evensen, 1994b]{Evensen1994a}
Evensen, G. (1994b).
\newblock Sequential data assimilation with a nonlinear quasi-geostrophic model
  using {Monte Carlo} methods to forecast error statistics.
\newblock {\em Journal of Geophysical Research}, 99(C5):10143.

\bibitem[Galli and Solidoro, 2018]{Galli2018}
Galli, G. and Solidoro, C. (2018).
\newblock {ATP} {Supply May Contribute to Light-Enhanced Calcification in
  Corals More Than Abiotic Mechanisms}.
\newblock {\em Frontiers in Marine Science}, 5.

\bibitem[Garbuno-Inigo et~al., 2020]{GarbunoInigo2020}
Garbuno-Inigo, A., Hoffmann, F., Li, W., and Stuart, A.~M. (2020).
\newblock {Interacting Langevin Diffusions: Gradient Structure and Ensemble
  Kalman Sampler}.
\newblock {\em {SIAM} Journal on Applied Dynamical Systems}, 19(1):412--441.

\bibitem[Girolami and Calderhead, 2011]{Girolami2011}
Girolami, M. and Calderhead, B. (2011).
\newblock {Riemann manifold Langevin and Hamiltonian Monte Carlo methods}.
\newblock {\em Journal of the Royal Statistical Society: Series B (Statistical
  Methodology)}, 73(2):123--214.

\bibitem[Gordon et~al., 1993]{Gordon1993}
Gordon, N.~J., Salmond, D.~J., and Smith, A. F.~M. (1993).
\newblock Novel approach to nonlinear/non-{G}aussian {B}ayesian state
  estimation.
\newblock {\em {IEE} Proceedings F Radar and Signal Processing}, 140(2):107.

\bibitem[Hastings, 1970]{Hastings1970}
Hastings, W.~K. (1970).
\newblock {Monte Carlo sampling methods using Markov chains and their
  applications}.
\newblock {\em Biometrika}, 57(1):97--109.

\bibitem[Holzworth et~al., 2014]{Holzworth2014}
Holzworth, D.~P., Huth, N.~I., deVoil, P.~G., Zurcher, E.~J., Herrmann, N.~I.,
  McLean, G., Chenu, K., van Oosterom, E.~J., Snow, V., Murphy, C., Moore,
  A.~D., Brown, H., Whish, J. P.~M., Verrall, S., Fainges, J., Bell, L.~W.,
  Peake, A.~S., Poulton, P.~L., Hochman, Z., Thorburn, P.~J., Gaydon, D.~S.,
  Dalgliesh, N.~P., Rodriguez, D., Cox, H., Chapman, S., Doherty, A., Teixeira,
  E., Sharp, J., Cichota, R., Vogeler, I., Li, F.~Y., Wang, E., Hammer, G.~L.,
  Robertson, M.~J., Dimes, J.~P., Whitbread, A.~M., Hunt, J., van Rees, H.,
  McClelland, T., Carberry, P.~S., Hargreaves, J. N.~G., MacLeod, N., McDonald,
  C., Harsdorf, J., Wedgwood, S., and Keating, B.~A. (2014).
\newblock {APSIM} {\textendash} evolution towards a new generation of
  agricultural systems simulation.
\newblock {\em Environmental Modelling {\&} Software}, 62:327--350.

\bibitem[Iglesias et~al., 2018]{Iglesias2018}
Iglesias, M., Park, M., and Tretyakov, M.~V. (2018).
\newblock Bayesian inversion in resin transfer molding.
\newblock {\em Inverse Problems}, 34(10):105002.

\bibitem[Iglesias and Yang, 2021]{Iglesias2021}
Iglesias, M. and Yang, Y. (2021).
\newblock Adaptive regularisation for ensemble {K}alman inversion.
\newblock {\em Inverse Problems}, 37(2):025008.

\bibitem[Iglesias, 2014]{Iglesias2014}
Iglesias, M.~A. (2014).
\newblock Iterative regularization for ensemble data assimilation in reservoir
  models.
\newblock {\em Computational Geosciences}, 19(1):177--212.

\bibitem[Iglesias et~al., 2013]{Iglesias2013}
Iglesias, M.~A., Law, K. J.~H., and Stuart, A.~M. (2013).
\newblock Ensemble {Kalman} methods for inverse problems.
\newblock {\em Inverse Problems}, 29(4):045001.

\bibitem[Jasra et~al., 2010]{Jasra2010}
Jasra, A., Stephens, D.~A., Doucet, A., and Tsagaris, T. (2010).
\newblock {Inference for L{\'{e}}vy-Driven Stochastic Volatility Models via
  Adaptive Sequential Monte Carlo}.
\newblock {\em Scandinavian Journal of Statistics}, 38(1):1--22.

\bibitem[Lan et~al., 2022]{Lan2022}
Lan, S., Li, S., and Shahbaba, B. (2022).
\newblock {Scaling Up Bayesian Uncertainty Quantification for Inverse Problems
  using Deep Neural Networks}.
\newblock {\em arXiv preprints}, page arXiv:2101.03906.

\bibitem[Le~Gland et~al., 2009]{LeGland2009}
Le~Gland, F., Monbet, V., and Tran, V.-D. (2009).
\newblock {Large sample asymptotics for the ensemble Kalman filter}.
\newblock Research Report RR-7014, {INRIA}.

\bibitem[Monsalve-Bravo et~al., 2022]{MonsalveBravo2022}
Monsalve-Bravo, G.~M., Lawson, B. A.~J., Drovandi, C., Burrage, K., Brown,
  K.~S., Baker, C.~M., Vollert, S.~A., Mengersen, K., McDonald-Madden, E., and
  Adams, M.~P. (2022).
\newblock Analysis of sloppiness in model simulations: unveiling parameter
  uncertainty when mathematical models are fitted to data.
\newblock {\em arXiv preprints}, page arXiv:2203.15184.

\bibitem[Probert et~al., 1998]{Probert1998}
Probert, M.~E., Dimes, J.~P., Keating, B.~A., Dalal, R.~C., and Strong, W.~M.
  (1998).
\newblock {APSIM}{\textquotesingle}s water and nitrogen modules and simulation
  of the dynamics of water and nitrogen in fallow systems.
\newblock {\em Agricultural Systems}, 56(1):1--28.

\bibitem[Rammay et~al., 2020]{Rammay2020}
Rammay, M.~H., Elsheikh, A.~H., and Chen, Y. (2020).
\newblock Flexible iterative ensemble smoother for calibration of perfect and
  imperfect models.
\newblock {\em Computational Geosciences}, 25(1):373--394.

\bibitem[Robert and Casella, 1999]{Robert1999}
Robert, C.~P. and Casella, G. (1999).
\newblock {\em {Monte Carlo Statistical Methods}}.
\newblock Springer New York.

\bibitem[Roberts and Rosenthal, 2001]{Roberts2001}
Roberts, G.~O. and Rosenthal, J.~S. (2001).
\newblock Optimal scaling for various {Metropolis-Hastings} algorithms.
\newblock {\em Statistical Science}, 16(4).

\bibitem[Rodolfo-Metalpa et~al., 2010]{RodolfoMetalpa2010}
Rodolfo-Metalpa, R., Martin, S., Ferrier-Pag\`es, C., and Gattuso, J.-P.
  (2010).
\newblock Response of the temperate coral {C}ladocora caespitosa to mid- and
  long-term exposure to p{CO}$_{2}$ and temperature levels projected for the
  year 2100 {AD}.
\newblock {\em Biogeosciences}, 7(1):289--300.

\bibitem[Roth et~al., 2017]{Roth2017}
Roth, M., Hendeby, G., Fritsche, C., and Gustafsson, F. (2017).
\newblock The {Ensemble} {Kalman} filter: a signal processing perspective.
\newblock {\em {EURASIP} Journal on Advances in Signal Processing}, 2017(1).

\bibitem[Sch{\"o}n and Lindsten, 2017]{Schoen2017}
Sch{\"o}n, T.~B. and Lindsten, F. (2017).
\newblock {Learning of dynamical systems--Particle filters and Markov chain
  methods}.
\newblock {\em Draft available}.

\bibitem[South et~al., 2019]{South2019}
South, L.~F., Pettitt, A.~N., and Drovandi, C.~C. (2019).
\newblock {Sequential Monte Carlo Samplers with Independent Markov Chain Monte
  Carlo Proposals}.
\newblock {\em Bayesian Analysis}, 14(3).

\bibitem[Vilas et~al., 2021]{Vilas2021}
Vilas, M., Bennett, F., Verburg, K., and Adams, M. (2021).
\newblock Considering unknown uncertainty in imperfect models: nitrogen
  mineralization as a case study.
\newblock In {\em {MODSIM}2021, 24th International Congress on Modelling and
  Simulation.}, pages 120--126. Modelling and Simulation Society of Australia
  and New Zealand.

\bibitem[Vollert et~al., 2022]{Vollert2022}
Vollert, S.~A., Drovandi, C., Monsalve-Bravo, G.~M., and Adams, M.~P. (2022).
\newblock Strategic model reduction by analysing model sloppiness: a case study
  in coral calcification.
\newblock {\em arXiv preprints}, page arXiv:2204.05602.

\bibitem[Wu et~al., 2022]{Wu2022}
Wu, J., Wen, L., Green, P.~L., Li, J., and Maskell, S. (2022).
\newblock {Ensemble Kalman filter based sequential Monte Carlo sampler for
  sequential Bayesian inference}.
\newblock {\em Statistics and Computing}, 32(1).

\end{thebibliography}

\appendix

\section{Extra results for the seagrass model} \label{A:seagrass}

\begin{figure}[H]
	\centering
	\includegraphics[scale=0.6]{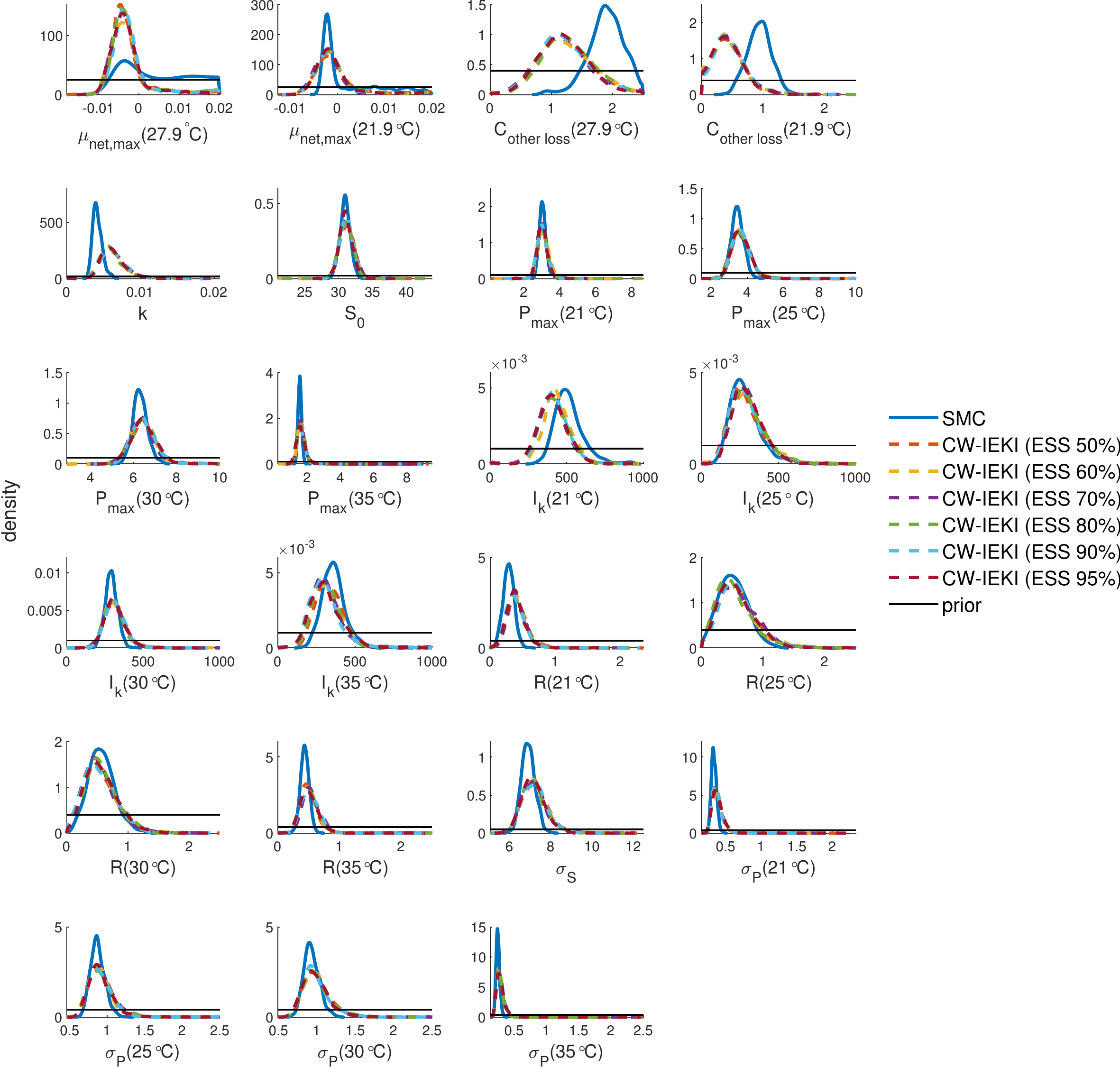}
	\caption{Marginal posterior density plots for the seagrass model applied to the \textit{H. uninervis} data.}
	\label{fig:Hu_densities}
\end{figure}

\begin{figure}[H]
	\hspace*{-2cm}    
	\centering
	\includegraphics[scale=0.6]{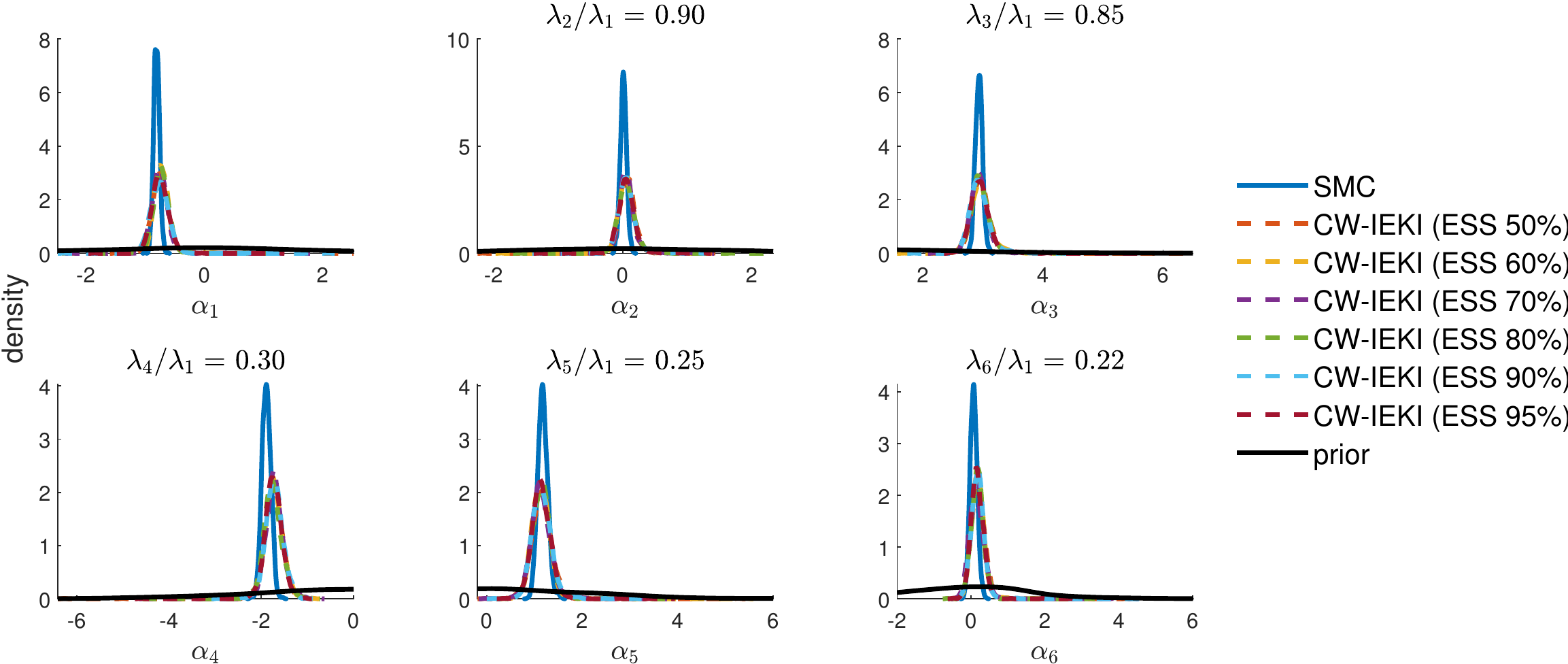} 
	\caption{Marginal density plots of the six stiffest eigenparameters (calculated using equation \eqref{eq:eigenparameters_seagrass}) for the seagrass model applied to the \textit{H. uninervis} data. Note that the uncertainty parameters in $\phi$ are excluded from the analysis of sloppiness, and $\lambda_k$ is the eigenvalue associated with eigenvector $v_k$ in equation \eqref{eq:eigenparameters_seagrass}. The eigenparameters are calculated based on samples from the prior (black), CW-IEKI (dashed) and SMC (blue) using a sensitivity matrix calculated using the SMC samples.}
	\label{fig:Hu_eigenparameter}
\end{figure}

\begin{figure}[H]
	\hspace*{-2cm}    
	\centering
	\includegraphics[scale=0.6]{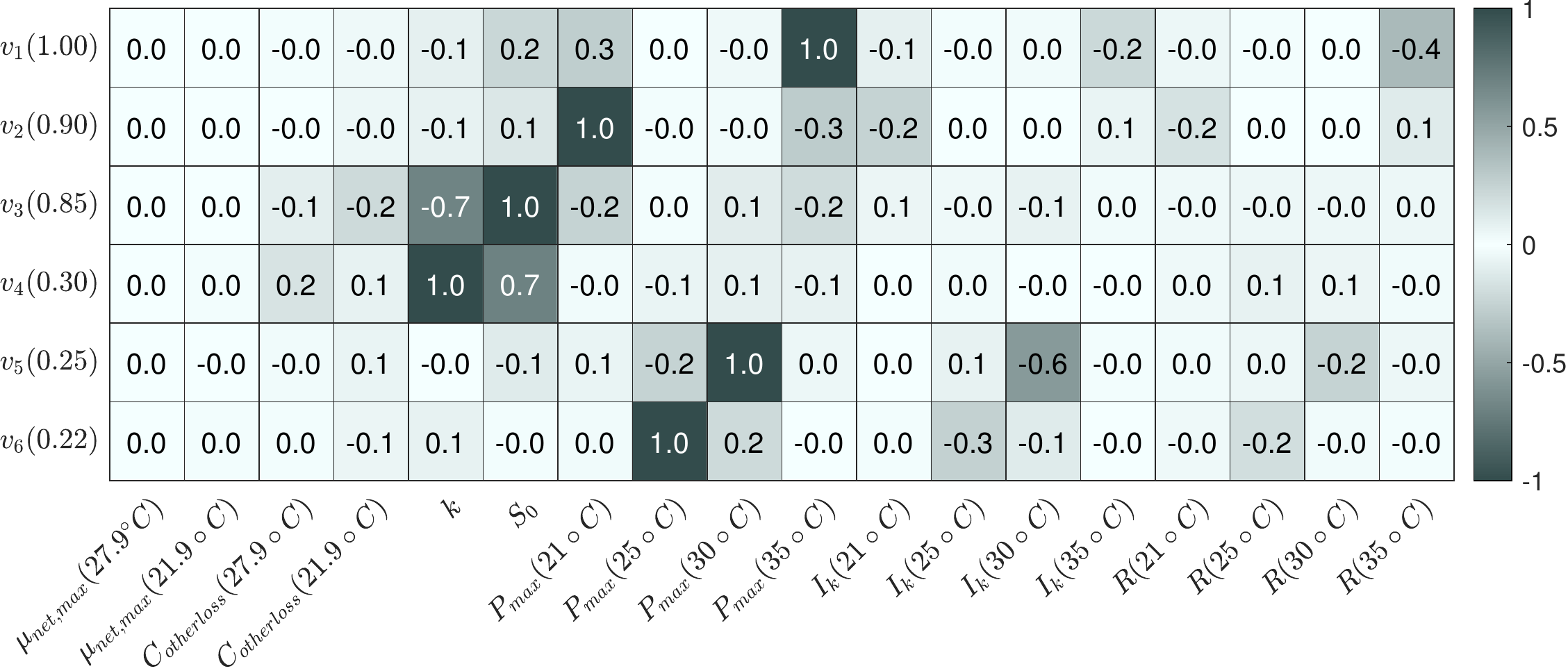} 
	\caption{Eigenvectors of the six stiffest eigenparameters for the seagrass model applied to the \textit{H. uninervis} data. These results are based on the SMC posterior samples. The labels on the left-hand side correspond to $v_k(\lambda_k\slash\lambda_1)$, where $v_k$ and $\lambda_k$ are the eigenvector and associated eigenvalue of eigenparameter $k$, respectively. The shade of the cells in row $k$ indicate the relative contribution ${(v_k)}_j$ of the $j$th parameter to eigenparameter $k$ --- parameters with darker colours have the greatest contribution.}
	\label{fig:Hu_eigenvectors}
\end{figure}

\begin{figure}[H]
	\hspace*{-2cm}    
	\centering
	\includegraphics[scale=0.6]{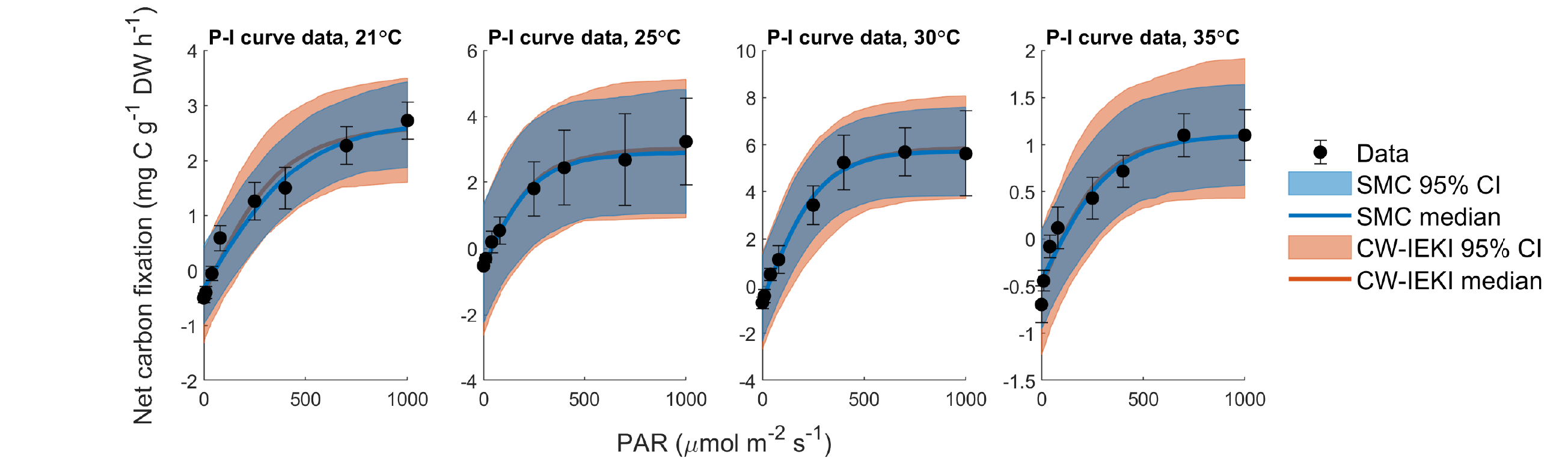}
	\caption{Comparison of the \textit{H. uninervis} data to the median and $95\%$ central credible intervals for the posterior predictive distribution of net carbon fixation obtained from the seagrass model fitted to this data. Models were fitted using CW-IEKI (red-orange) and likelihood tempering SMC (blue). P-I $=$ photosynthesis-irradiance and PAR $=$ photosynthetically active radiation \citep[see][]{Adams2020}.}
	\label{fig:Hu_postpred_PI}
\end{figure}

\begin{figure}[H]
	\hspace*{-2cm}    
	\centering
	\includegraphics[scale=0.6]{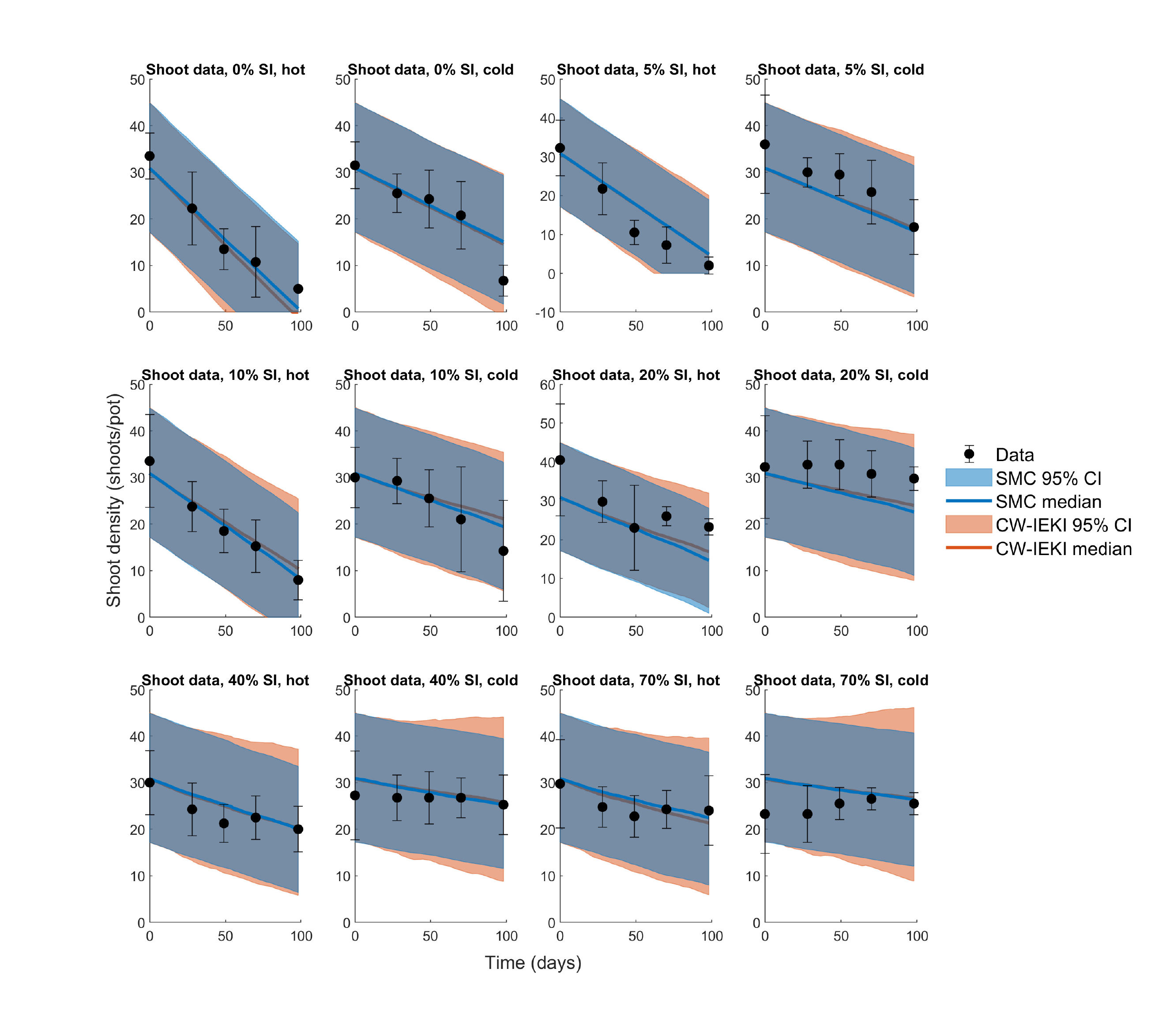}
	\caption{Comparison of the \textit{H. uninervis} data to the median and $95\%$ central credible intervals for the posterior predictive distribution of shoot density obtained from the seagrass model fitted to this data. Models were fitted using CW-IEKI (red-orange) and likelihood tempering SMC (blue). SI $=$ surface irradiance and ``hot" or ``cold" indicates the temperature conditions under which the seagrass data was collected \citep{Adams2020}.}
	\label{fig:Hu_postpred_SD}
\end{figure}

\begin{table}[H]
	\centering
	\begin{tabular}{|c|c|cccccc|}
		\hline
		Method & SMC & \multicolumn{6}{c|}{CW-IEKI} \\
		ESS target threshold & $50\%$ & $95\%$ & $90\%$ & $80\%$ & $70\%$ & $60\%$ & $50\%$ \\
		\hline
		$G(\cdot)$ evaluations & 594000 & 86000 & 58000 & 38000 & 29000 & 24000 & 20000 \\
		Approximate speed-up & 1.00 & 6.91 & 10.24 & 15.63 & 20.48 & 24.75 & 29.70 \\
		\hline 
	\end{tabular}
	\caption{Total and relative number of evaluations of $G(\cdot)$ for SMC and CW-IEKI with different ESS target thresholds. Results are for the seagrass model applied to the \textit{H. uninervis} data. Note that the total number of evaluations of $G(\cdot)$ is a multiple of the number of samples $N = 1000$.}
\end{table}

\begin{figure}[H]
	\centering
	\includegraphics[scale=0.6]{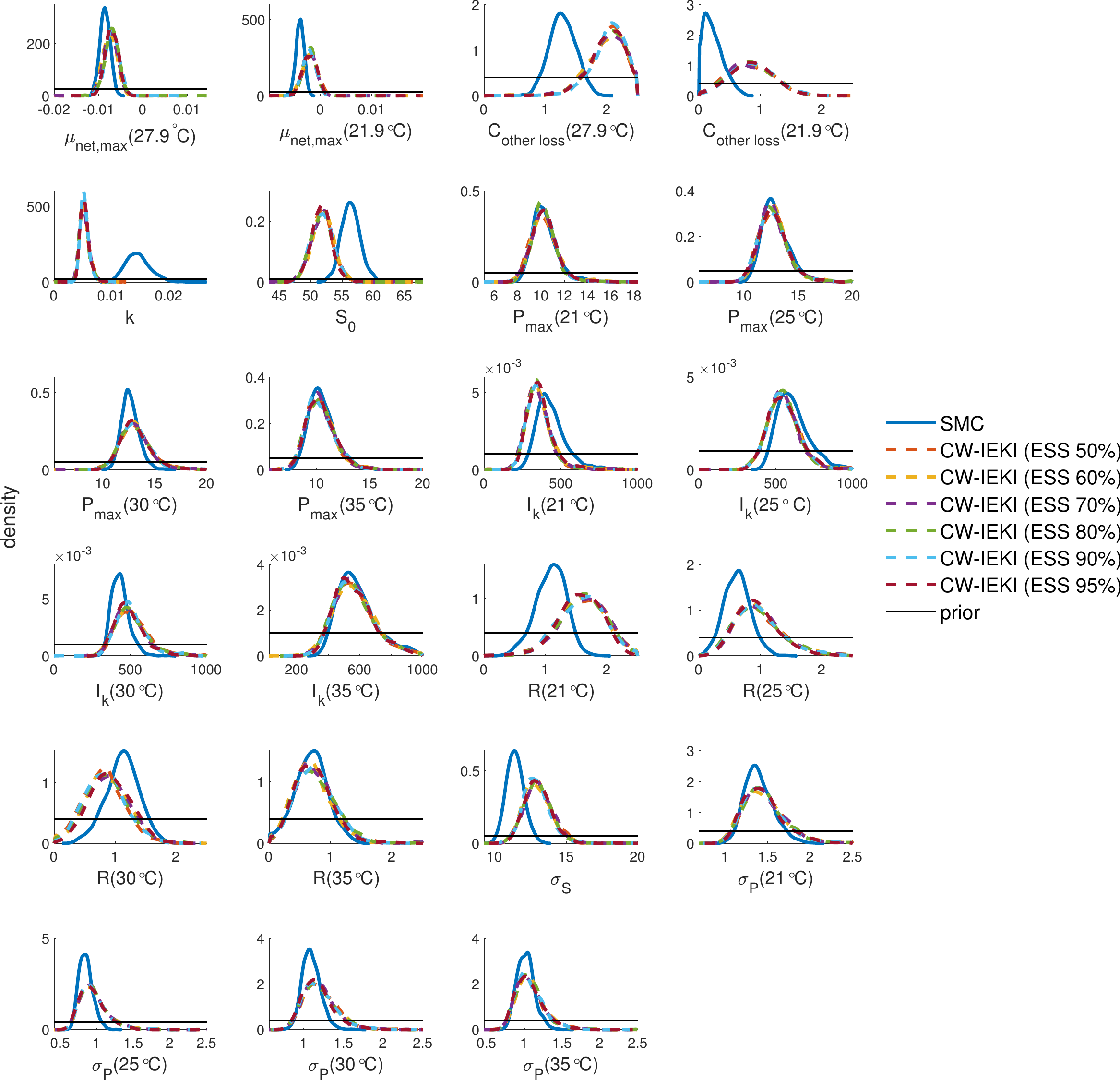}
	\caption{Marginal posterior density plots for the seagrass model applied to the \textit{Z. muelleri} data.}
	\label{fig:Zm_densities}
\end{figure}

\begin{figure}[H]
	\hspace*{-2cm}    
	\centering
	\includegraphics[scale=0.6]{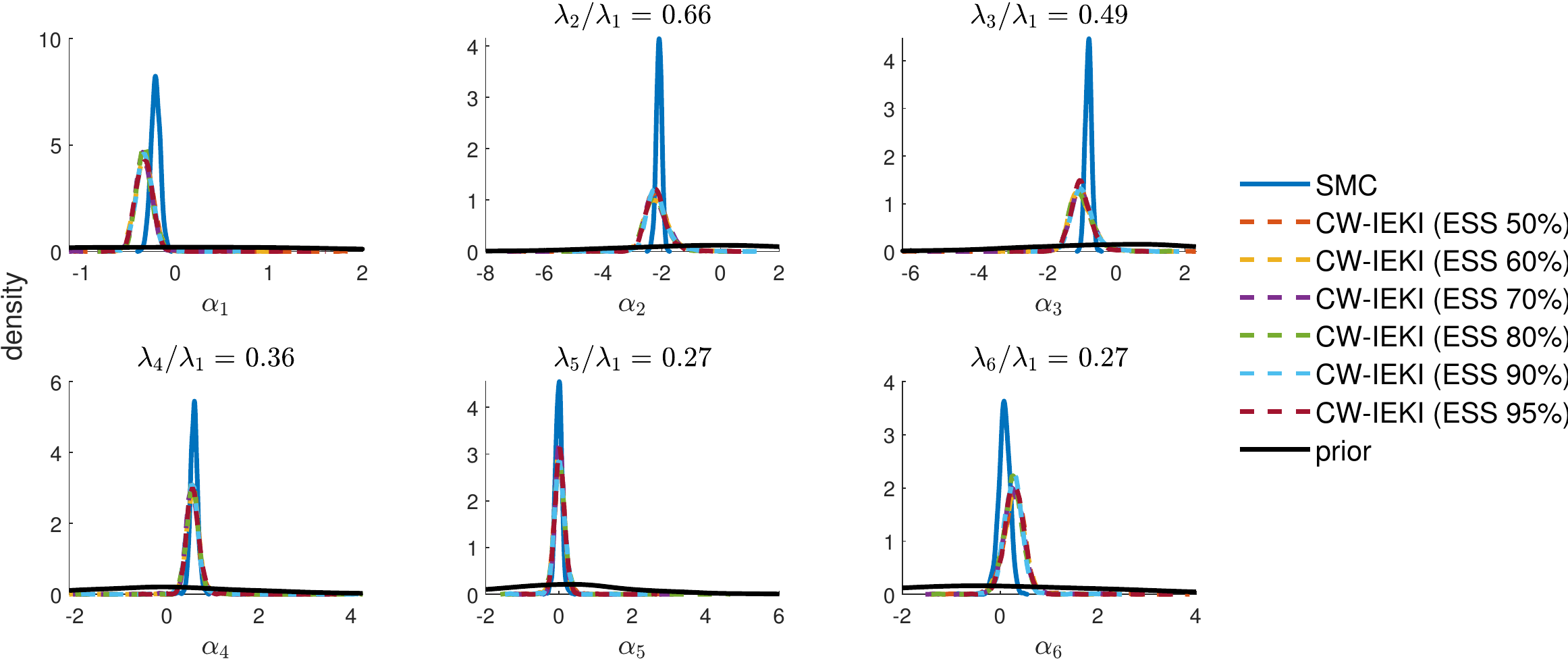} 
	\caption{Marginal density plots of the six stiffest eigenparameters (calculated using equation \eqref{eq:eigenparameters_seagrass}) for the seagrass model applied to the \textit{Z. muelleri} data. Note that the uncertainty parameters in $\phi$ are excluded from the analysis of sloppiness, and $\lambda_k$ is the eigenvalue associated with eigenvector $v_k$ in equation \eqref{eq:eigenparameters_seagrass}. The eigenparameters are calculated based on samples from the prior (black), CW-IEKI (dashed) and SMC (blue) using a sensitivity matrix calculated using the SMC samples.}
	\label{fig:Zm_eigenparameter}
\end{figure}

\begin{figure}[H]
	\hspace*{-2cm}    
	\centering
	\includegraphics[scale=0.6]{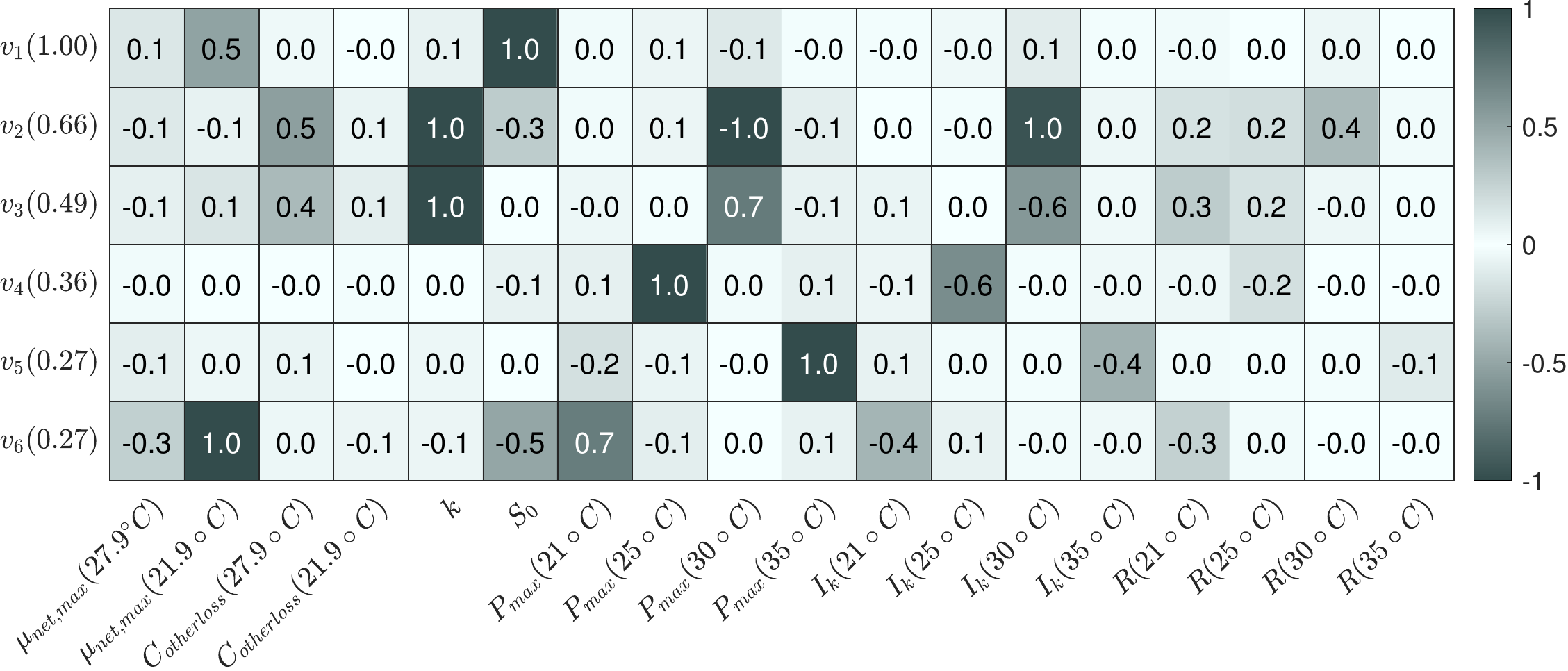} 
	\caption{Eigenvectors of the six stiffest eigenparameters for the seagrass model applied to the \textit{Z. muelleri} data. These results are based on the SMC posterior samples. The labels on the left-hand side correspond to $v_k(\lambda_k\slash\lambda_1)$, where $v_k$ and $\lambda_k$ are the eigenvector and associated eigenvalue of eigenparameter $k$, respectively. The shade of the cells in row $k$ indicate the relative contribution ${(v_k)}_j$ of the $j$th parameter to eigenparameter $k$ --- parameters with darker colours have the greatest contribution.}
	\label{fig:Zm_eigenvectors}
\end{figure}

\begin{figure}[H]
	\hspace*{-2cm}    
	\centering
	\includegraphics[scale=0.6]{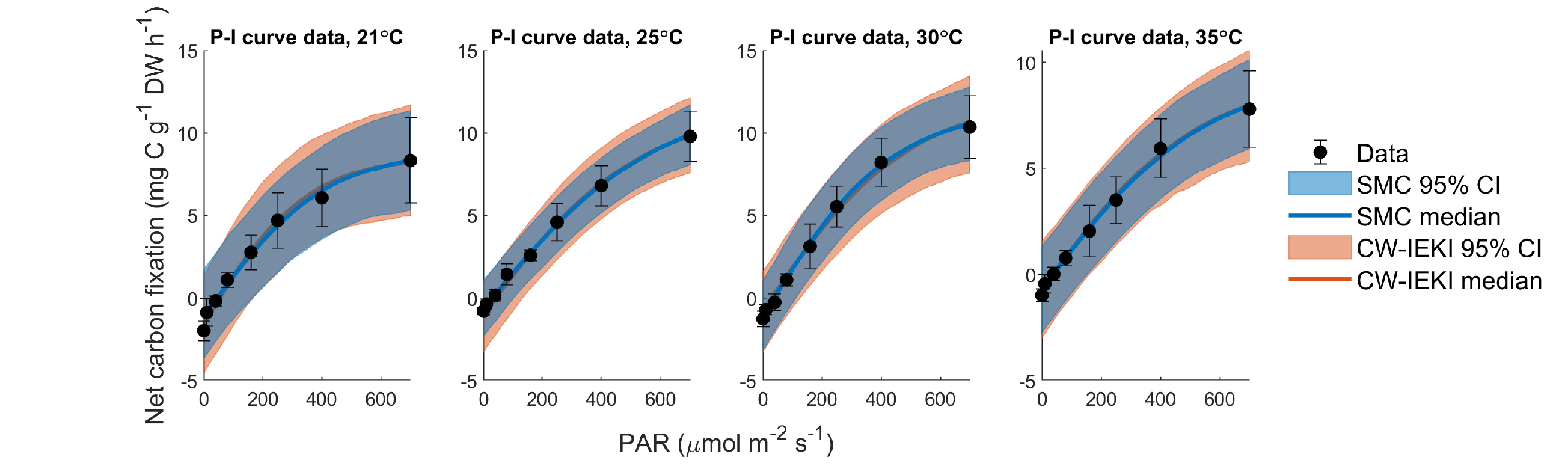}
	\caption{Comparison of the \textit{Z. muelleri} data to the median and $95\%$ central credible intervals for the posterior predictive distribution of net carbon fixation obtained from the seagrass model fitted to this data. Models were fitted using CW-IEKI (red-orange) and likelihood tempering SMC (blue). P-I $=$ photosynthesis-irradiance and PAR $=$ photosynthetically active radiation \citep[see][]{Adams2020}.}
	\label{fig:Zm_postpred_PI}
\end{figure}

\begin{figure}[H]
	\hspace*{-2cm}    
	\centering
	\includegraphics[scale=0.6]{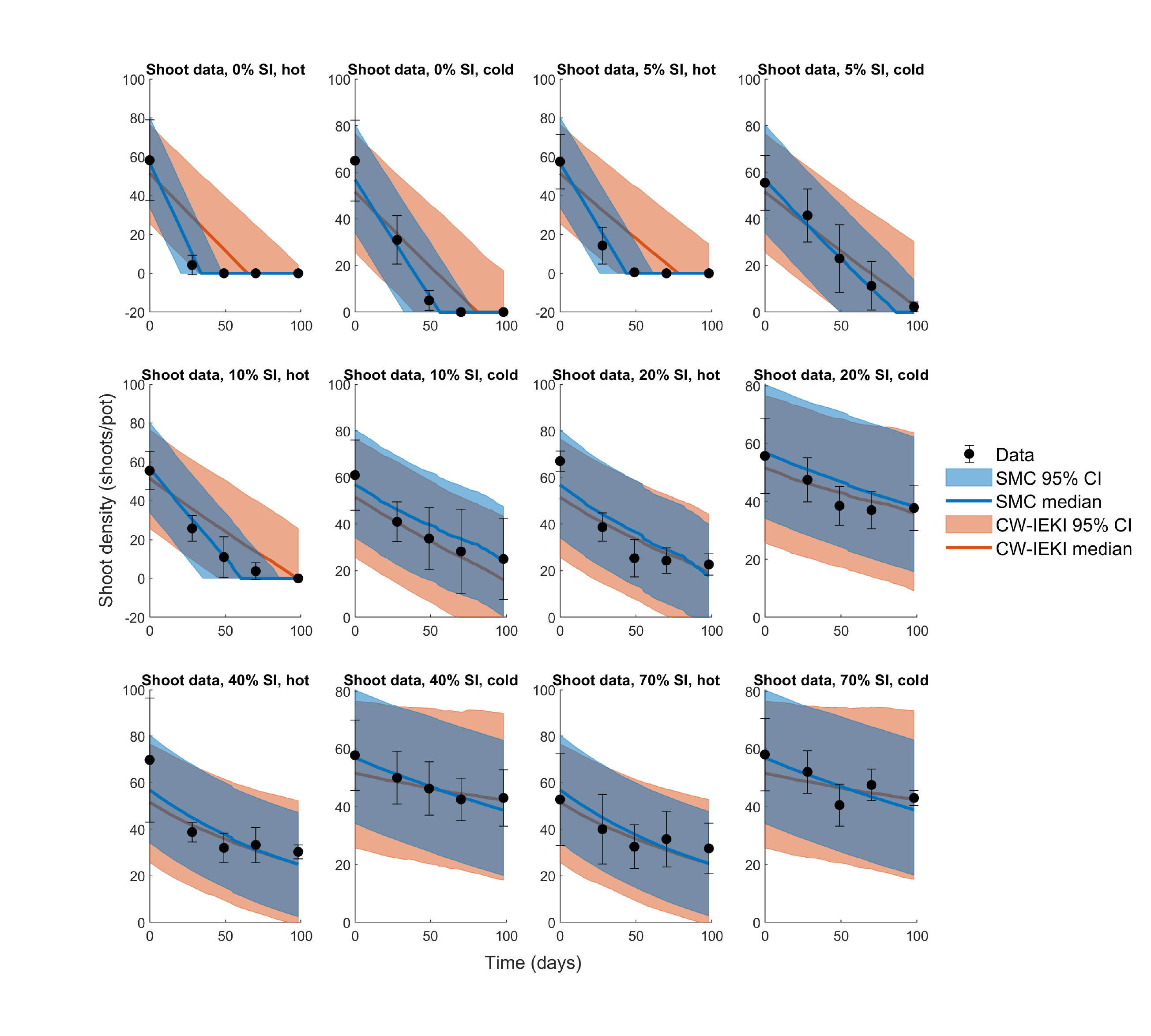}
	\caption{Comparison of the \textit{Z. muelleri} data to the median and $95\%$ central credible intervals for the posterior predictive distribution of shoot density obtained from the seagrass model fitted to this data. Models were fitted using CW-IEKI (red-orange) and likelihood tempering SMC (blue). SI $=$ surface irradiance and ``hot" or ``cold" indicates the temperature conditions under which the seagrass data was collected \citep{Adams2020}.}
	\label{fig:Zm_postpred_SD}
\end{figure}

\begin{table}[H]
	\centering
	\begin{tabular}{|c|c|cccccc|}
		\hline
		Method & SMC & \multicolumn{6}{c|}{CW-IEKI} \\
		ESS target threshold & $50\%$ & $95\%$ & $90\%$ & $80\%$ & $70\%$ & $60\%$ & $50\%$ \\
		\hline
		$G(\cdot)$ evaluations & 686000 & 68000 & 46000 & 30000 & 23000 & 19000 & 16000 \\
		Approximate speed-up & 1.00 & 10.09 & 14.91 & 22.87 & 29.83 & 36.11 & 42.88 \\
		\hline 
	\end{tabular}
	\caption{Total and relative number of evaluations of $G(\cdot)$ for SMC and CW-IEKI with different ESS target thresholds. Results are for the seagrass model applied to the \textit{Z. muelleri} data. Note that the total number of evaluations of $G(\cdot)$ is a multiple of the number of samples $N = 1000$.}
\end{table}

\section{Extra results for the coral model} \label{A:coral}

\begin{figure}[H]
	\centering
	\includegraphics[scale=0.7]{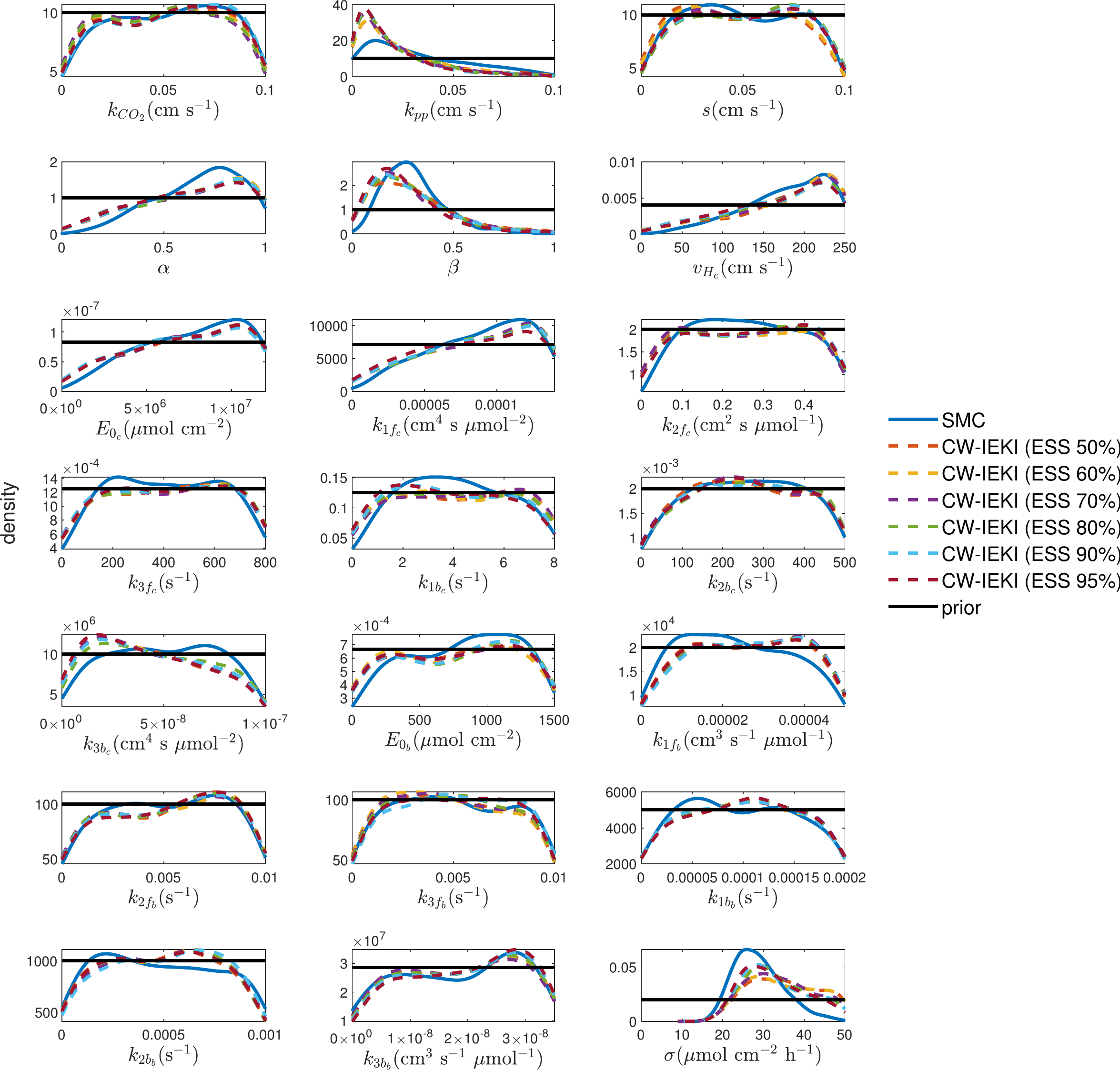}
	\caption{Marginal posterior density plots for the coral model.}
	\label{fig:coral_densities}
\end{figure}

\end{document}